\documentclass[fleqn,usenatbib]{mnras}

\usepackage{newtxtext,newtxmath}
\usepackage{comment}
\usepackage[T1]{fontenc}
\usepackage{graphicx}	
\usepackage{amsmath}	
\usepackage{xcolor}

\DeclareRobustCommand{\VAN}[3]{#2}
\let\VANthebibliography\thebibliography
\def\thebibliography{\DeclareRobustCommand{\VAN}[3]{##3}\VANthebibliography}

\title[Dust evolution in a cosmological simulation]{Evolution of the grain size distribution in Milky Way-like galaxies in post-processed IllustrisTNG simulations}

\author[Y.-H. Huang et al.]{Yu-Hsiu Huang$^{1,2}$\thanks{E-mail: yuhsiuhuang@asiaa.sinica.edu.tw}, 
Hiroyuki Hirashita$^{2}$, 
Yun-Hsin Hsu$^{2,3}$,
Yen-Ting Lin$^{2}$, 
\newauthor Dylan Nelson$^{4}$,
Andrew P. Cooper$^{3,5,6}$
\\
$^{1}$Institute of Physics, National Taiwan University, No.\ 1, Section 4, Roosevelt Road, Taipei 10617, Taiwan\\
$^{2}$Institute of Astronomy and Astrophysics, Academia Sinica,
Astronomy-Mathematics Building, No.\ 1, Section 4,\\
Roosevelt Road, Taipei 10617, Taiwan \\
$^{3}$Institute of Astronomy, National Tsing Hua University, 101, Section 2, Kuang-Fu Road, Hsinchu 30014, Taiwan\\
$^{4}$Universit\"{a}t Heidelberg, Zentrum f\"{u}r Astronomie, Institut f\"{u}r theoretische Astrophysik, Albert-Ueberle-Str. 2, 69120 Heidelberg, Germany\\
$^{5}$Department of Physics, National Tsing Hua University, 101, Section 2, Kuang-Fu Road, Hsinchu 30014, Taiwan\\
$^{6}$Center for Informatics and Computation in Astronomy, National Tsing Hua University, 101, Section 2, Kuang-Fu Road, Hsinchu 30013, Taiwan
}

\date{}
\pubyear{2020}

\begin{document}
\label{firstpage}
\pagerange{\pageref{firstpage}--\pageref{lastpage}}
\maketitle

\begin{abstract}
We model dust evolution in Milky Way-like galaxies by post-processing the IllustrisTNG cosmological hydrodynamical simulations in order to predict dust-to-gas ratios and grain size distributions. We treat grain-size-dependent dust growth and destruction processes using a 64-bin discrete grain size evolution model without spatially  resolving  each galaxy. Our model broadly reproduces the observed dust--metallicity scaling relation in nearby galaxies. The grain size distribution is dominated by large grains at $z\gtrsim 3$ and the small-grain abundance rapidly increases by shattering and accretion (dust growth) at $z\lesssim 2$. The grain size distribution approaches the so-called MRN distribution at $z\sim 1$, but a suppression of large-grain abundances occurs at $z<1$. Based on the computed grain size distributions and grain compositions, we also calculate the evolution of the extinction curve for each Milky Way analogue. Extinction curves are initially flat at $z>2$, and become consistent with the Milky Way extinction curve at $z\lesssim 1$ at $1/\lambda < 6~\micron^{-1}$. However, typical extinction curves predicted by our model have a steeper slope at short wavelengths than is observed in the Milky Way. This is due to the low-redshift decline of gas-phase metallicity and the dense gas fraction in our TNG Milky Way analogues that suppresses the formation of large grains through coagulation.
\end{abstract}



\section{Introduction} \label{sec:intro}

Although dust only accounts for a small fraction of the mass of the interstellar medium (ISM), $\sim$\,1 percent in the case of the Milky Way (MW), it plays an important role in several aspects of galaxy evolution. Dust absorbs the ultraviolet (UV) radiation from stars and re-emits the energy in the far-infrared \citep[FIR, e.g.][]{Takeuchi:2005mnras}. In this way, the spectral energy distributions (SEDs) of galaxies are modified by dust. Thus, correcting for dust extinction is crucial to estimate the total intrinsic stellar luminosity and the star formation rate (SFR) of galaxies \citep[e.g.][]{Buat:1996aa, Inoue:2000pasj}. 
In addition to its influence on observables, dust is an efficient catalyst for the formation of molecular hydrogen \citep[H$_2$;][]{Gloud63, Cazaux04}, which is the main component in star-forming molecular clouds. 
Dust also locks gas-phase metals into the solid phase, thus serving as a depletion source of elements \citep[e.g.][]{Sauvage:1996araa, Jenkins:2009apj}. This not only affects the measurement of metallicity but also reduces the physical effects of metals, such as metal line cooling.

The radiative and chemical effects of dust are regulated not only by the total dust abundance, but also by the grain size distribution. The scattering and absorption of light by a dust grain depends on the grain size. The wavelength dependence of extinction (scattering and absorption), which is referred to as the extinction curve, depends on the grain size distribution. Indeed, \citet[][hereafter MRN]{Mathis:1977apj} extracted the grain size distributions of silicate and graphite from the observed MW extinction curve. In general, short-wavelength electromagnetic waves are more likely to be scattered or absorbed by small grains \citep{Bohren:1983book}. Thus, steeper extinction curves are observed if small grains dominate. The H$_2$ formation rate also depends on the grain size distribution, which governs the total grain surface area \citep{Yamasawa:2011apj}.

In order to predict the evolution of dust abundance and grain size distribution in galaxies, we need to treat dust enrichment and relevant physical processes in the ISM. A comprehensive model for the evolution of dust abundance and grain size distribution can be found in \citet{Asano:2013aa}. There are two major sources of dust formation: dust condensation in stellar winds and supernova (SN) ejecta, and dust growth through the accretion of gas-phase metals. Dust is also destroyed by SN shocks sweeping through the ISM. There are two grain--grain collision effects that change grain sizes while approximately conserving total dust mass: coagulation (grain--grain sticking) and shattering (grain disruption). Of these five processes, stellar dust production plays the most significant role in the early epoch of galaxy evolution, because the remaining four processes require the presence of either metals or dust grains. At later times, when a galaxy contains a sufficient amount of metals and dust, shattering and accretion work together to increase the small-grain population while coagulation increases the large-grain population. Meanwhile, SN destruction continues to destroy grains, which is balanced by the major dust production mechanisms (stellar dust production and accretion at early and late stages, respectively). 

The above processes, especially those concerning dust processing, are dependent on the environmental properties of the ISM (gas density, gas temperature, metallicity, etc.). 
Thus, we expect that the grain size distribution varies from galaxy to galaxy and with the galaxy age. 
Hydrodynamical simulations are one means of self-consistently modelling the evolution of dust alongside the other properties of the ISM. In particular, state-of-the-art cosmological simulations can now trace the independent evolution of metals, stars and the multi-phase interstellar medium on kiloparsec scales \citep[e.g.][]{Pillepich18b, Shimizu:2019mnras}. The physical processes described above can be used to calculate how dust evolves, consistently with the varying physical conditions of the ISM.

However, it is computationally expensive to implement detailed dust evolution prescriptions in galaxy-scale or cosmological simulations. For this reason, most simulations model the effect of dust by post-processing the ISM predicted by existing simulations, making simple assumptions about dust abundance and properties \citep[e.g.][]{Yajima:2015mnras, Ma19, Vogelsberger20, Trcka20}. Despite the computational difficulties, several cosmological simulations have attempted to include an in-situ treatment of dust \citep{Bekki:2015mnras, Aoyama:2018mnras, Hou:2019aa,Graziani2020} in different ways including zoom-in simulations coupled with dust process models \citep{McKinnon16} and cosmological-volume simulations with single-size dust evolution models \citep{McKinnon17, Dave19}. 
In addition to cosmological simulations, some studies focus on isolated galaxies  \citep{Bekki15:apj, Aoyama:2017mnras, Hou:2017mnras, McKinnon18, Aoyama:2020mnras} or on a sub-region within a galaxy \citep{Zhukovska:2016apj, Hu:2019mnras}. In such simulations, it is possible to include the evolution of the grain size distribution either with a two-size approximation which follows only two representative grain sizes \citep{Aoyama:2017mnras, Hou:2017mnras}, or with full-size modelling by resolving the entire grain size distribution \citep{McKinnon18, Aoyama:2020mnras}.
The two-size approximation has also been implemented in zoom-in cosmological simulations \citep{Gjergo2018,Granato2020}.
These simulations emphasize the importance of capturing physical ISM conditions in order to evolve the grain size distribution.

\citet{Aoyama:2018mnras} and \citet{Hou:2019aa} adopted the two-size approximation to trace dust evolution in their cosmological simulations. The inclusion of grain size information enabled them to include more processes such as shattering and coagulation, and to predict extinction curves for which the grain size distribution is of fundamental importance. However, the two-size approximation has a limitation in that it has only two degrees of freedom in extinction curves. Thus, the precision and variety in the predicted extinction curves are limited. Similarly, evolving dust abundances could also be affected by the treatment of grain size distribution.
Recently, \citet{Aoyama:2020mnras} implemented the evolution of a full grain size distribution in a simulation of an isolated spiral galaxy. They predicted a tighter dispersion in the relation between dust-to-gas ratio and metallicity at low metallicity than  \citet{Aoyama:2017mnras}, who adopted the two-size approximation. This update arose from the difference in the efficiency of dust growth; that is, the two-size approximation may overestimate the rate of dust growth by accretion in low-metallicity environments. Other treatments of grain size distributions, such as the method of moments \citep{Mattsson16}, provide an alternative computational method, but would not improve the situation drastically as far as the balance between computational cost and precision is concerned. To overcome the limitations and uncertainties in simplified approximations including the two-size model, an implementation of the full grain size distribution is ideal.

In this paper, we make a first attempt to calculate the evolution of the full grain size distribution, resolving 64 bins in grain radii, using a large-volume cosmological simulation. In lieu of an on-the-fly implementation, we calculate the dust evolution by post-processing an existing simulation. Thus, we first describe the scheme we have developed for calculating the grain size distribution based on the output of cosmological simulations. We choose \textit{The Next Generation Illustris} project (hereafter IllustrisTNG or TNG, see more information about TNG in Section \ref{model:TNG}) as the model input in this work. In particular we use the TNG300-1 simulation, one of the TNG simulations, which combines detailed, well-constrained ISM predictions with a statistically representative cosmological volume. 

Our second goal is to compare the output of our model against observational constraints on dust properties. As shown by \citet{ODonnell97} and \citet{Asano14} using the MW extinction curve as a standard reference, extinction curves strongly reflect the evolution of the grain size distribution. We also focus on MW-like galaxies and compare our results with the MW extinction curve. In our previous studies using a one-zone model, \citet[][hereafter HM20]{2020DustModel} showed that MW-like extinction curves can be reproduced, although the ISM evolution was not taken into account. Our work in this paper builds on the HM20 model in a form suitable for post-processing cosmological simulations. This gives a first look at how the grain size distribution evolves in response to changes in the ISM across plausible galactic assembly histories. We can therefore examine whether the varied assembly histories of MW-like galaxies can produce dust properties and extinction curves comparable to those observed in the MW.
Some studies, using a semi-analytic treatment for the assembly of the MW, reproduced the dust mass in the MW \citep{deBennassuti2014,Ginolfi2018}. In this work, we predict the grain size distribution for the MW-like galaxies for the first time.

For extragalactic objects such as nearby spiral galaxies, it is difficult to observationally derive extinction curves. Thus, the MW extinction curve is a unique observable that reflects the optical properties of dust in MW-like galaxies, under the assumption that the MW is a typical $\mathrm{M^*}$ disk galaxy. For extragalactic objects it is easier to obtain \textit{attenuation} curves, which are effective extinction curves including radiation transfer effects within a galaxy \citep{Calzetti2001}. However, predicting attenuation properties requires additional information such as the star--dust distribution geometry and age-dependent stellar distribution \citep{Granato00, Panuzzo07, Walcher11, Salmon16, Narayanan18, Salim20}, and relies on proper radiative transfer calculations \citep[e.g.][]{Witt96, Pierini04, Narayanan18, Trayford20}. As shown by previous studies, attenuation curves are strongly modified by these effects. As mentioned later, the model we develop in this paper is not suitable for calculating attenuation curves because it does not take into account the detailed spatial distribution of gas and stars within a galaxy. Thus, we concentrate on extinction curves and leave detailed radiative transfer calculations for a future work. 

This paper is organized as follows. In Section \ref{sec:model} we review the cosmological simulation we adopt, the dust evolution model we use, and modifications made for post-processing. We show our results in Section \ref{sec:results} and discuss them in Section \ref{sec:discussion}. Our conclusions are summarized in Section \ref{sec:conclusion}. We adopt the solar metallicity $Z_\odot = 0.0127$ and the same cosmology as TNG: $h=0.6774$, $\Omega_\Lambda=0.6911$, $\Omega_m=0.3089$, and $\Omega_b=0.0486$ \citep{Planck16}.

\section{Methods and Dust Modelling} \label{sec:model}

In this section, we briefly review the IllustrisTNG cosmological simulation used in this work, and describe our post-processing model for dust evolution. In a separate subsection we focus on the major new component of our model -- the treatment of galaxy mergers.

\subsection{IllustrisTNG and the Milky Way sample selection} \label{model:TNG}
The IllustrisTNG project \citep{Pillepich18b, Springel18, Nelson18, Naiman18, Marinacci18} is a state-of-the-art cosmological galaxy formation simulation suite of cubic volumes with comoving side lengths of approximately 50, 100, and 300 Mpc (TNG50, TNG100, and TNG300, respectively). The spatial resolution is higher in the smaller volumes. Each simulation solves for the coupled evolution of dark matter and baryons from the early Universe to the present day, and includes comprehensive baryonic physics and feedback models \citep{Weinberger17,Pillepich18a}. The results of TNG100 and TNG300 are publicly available \citep{Nelson19}\footnote{\url{https://www.tng-project.org}}, along with halo/subhalo catalogs, merger trees, and other supplementary data.

TNG identifies virialized overdensities (haloes) and their self-bound internal structures (subhaloes) using the Friend-of-Friend \citep[FoF;][]{Davis85} and \textsc{Subfind} algorithms \citep{Springel01}. Galaxies are identified as the baryonic matter bound to haloes and subhaloes.

In this first application of our post-processing scheme we treat each TNG galaxy as a one-zone object. For this purpose, the number of sample galaxies (i.e. total volume) is more important than the spatial resolution. Thus, we adopt the TNG300-1 simulation as our primary tool, with a volume of $302.6^3~\mathrm{Mpc^3}$ and a baryonic mass resolution of $7.6\times10^6~\mathrm{M}_{\sun} h^{-1}$. See Table 1 of \citet{Nelson19} for detailed information about the parameters of TNG300-1. 

\begin{figure*}
    \centering
    \includegraphics[width=\textwidth]{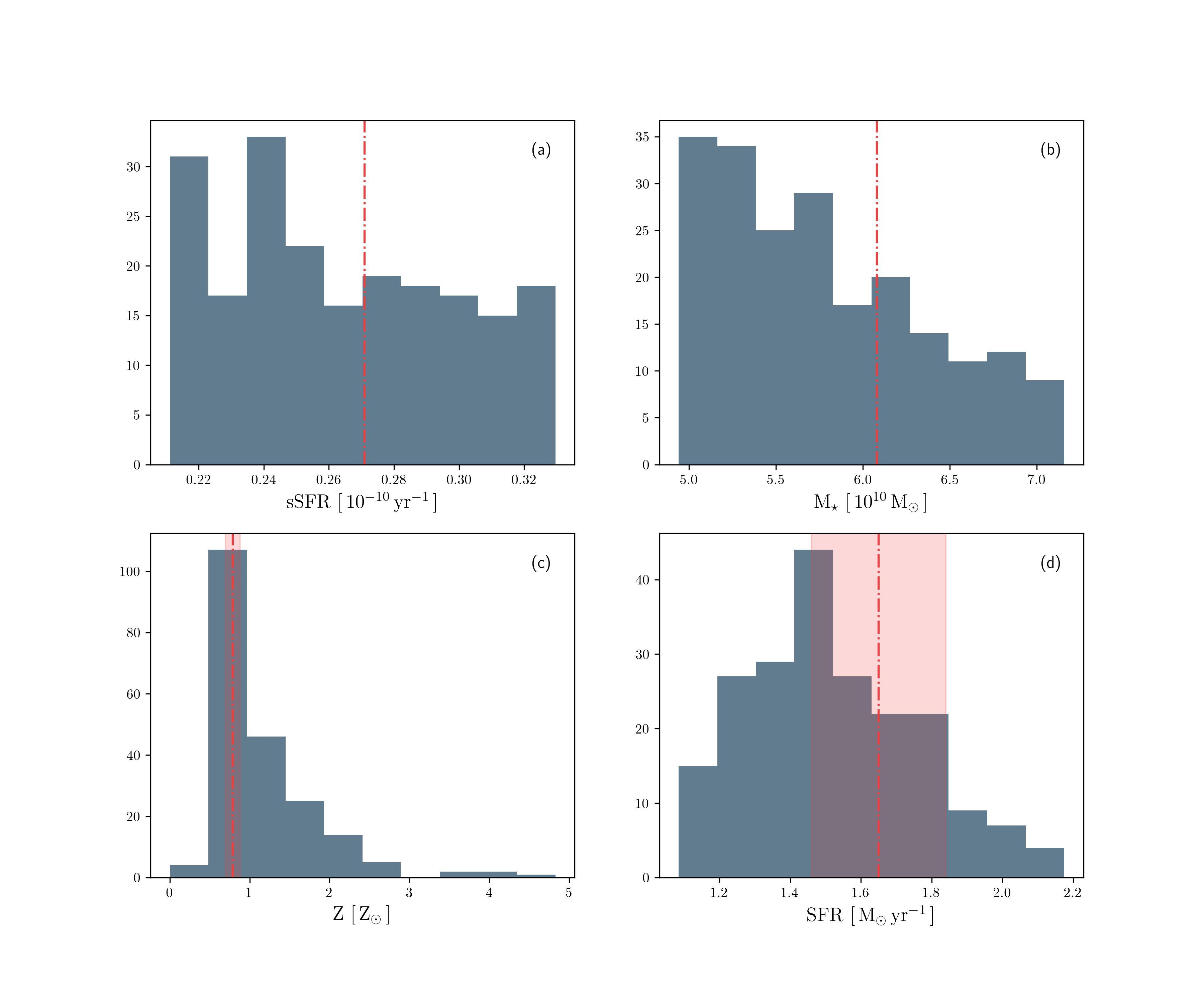}
    \caption{Distributions of (a) sSFR, (b) stellar mass, (c) metallicity, and (d) SFR of the 210 MW-like galaxies selected by the criteria we apply to TNG300 (see text). The histogram shows the number of galaxies in each of 10 equal-width bins. The red dash-dotted lines indicate the observed values for the MW and the shaded areas are the 1-$\sigma$ uncertainty \citep{Andre:2003apj, MW_like_def}. We do not show the uncertainty of sSFR and $M_*$ since we directly use them to select the sample.}
    \label{fig:sample}
\end{figure*}

To focus on MW-like galaxies we first define a sample of MW analogues in TNG by selecting on stellar mass $M_*$ and specific SFR (sSFR $= \mathrm{SFR}/M_*$). \citet{MW_like_def} estimate global properties of the MW using a hierarchical Bayesian statistical method. They constrain the MW stellar mass to be $M_*=6.08\pm 1.14\times10^{10}\,\mathrm{M_\odot}$ with $\mathrm{sSFR}=2.71\pm0.59\times10^{-11}\,\mathrm{yr^{-1}}$. We sample the 1-$\sigma$ dispersion for our selection, yielding 210 galaxies in TNG300 at redshift ($z$) zero.

The distributions of sSFR and $M_*$ for this sample are shown in Figs.\ \ref{fig:sample}a and b, respectively. We observe that the sSFR distribution is quite uniform whereas the stellar mass distribution is slightly biased towards the lower end of the mass range, as expected given the volume-limited parent sample. To contrast our sample against properties of the real Milky Way we show the distribution of metallicity and absolute SFR in Figs.\ \ref{fig:sample}c and d, respectively. The metallicities of galaxies in our sample are concentrated around $Z\sim 1\,Z_\odot$, as expected for MW-like galaxies (shown in red), although a few galaxies have $z\gtrsim 4\,Z_\odot$. The SFR is mostly in the range of 1--2 $\mathrm{M_\odot\,yr^{-1}}$ with a median of $\simeq 1.5\,\mathrm{M_\odot\,yr^{-1}}$, as compared to the $\simeq 1.6\,\mathrm{M_\odot\,yr^{-1}}$ inferred for the Milky Way (also shown in red).

TNG provides the merger history of each galaxy, i.e. merger trees, and we use the \textsc{SubLink} merger tree \citep{Rodriguez-Gomez15}. The link from a galaxy to its most massive progenitor at each earlier simulation snapshot defines the main branch of the merger tree. In this work, we apply the dust evolution model to every subhalo in a given merger tree, but we only present the evolution of size distributions and dust properties along the main branch.

For the dust evolution calculation described in Section \ref{model:size}, we take the gas mass, the stellar mass, the gas metallicity $Z$, the silicon and carbon abundances ($Z_\mathrm{Si}$ and $Z_\mathrm{C}$, respectively), and the SFR from TNG as inputs. Our fiducial choice is to adopt the gas mass within twice the stellar half mass radius, denoted as $M_\mathrm{gas,2R*}$. As described in Section \ref{combined:merger}, the gas masses of merging progenitors are used to determined the resulting dust-to-gas ratio after the merger. We are interested in interstellar (rather than circumgalactic) dust, which is best represented by this definition (i.e. aperture). Nevertheless, because the extent of the `interstellar' gas is ambiguous, we also examine the total gas mass $M_\mathrm{gas,tot}$ in the galaxy as a more extreme choice.
    
\subsection{Evolution of grain size distribution} \label{model:size}

Our model for the evolution of dust in galaxies is based on our previous HM20 treatment, itself based on a model for the grain size distribution primarily formulated by \citet{Asano:2013aa} and \cite{2019DustModel}. We briefly review the model and refer the interested reader to HM20 for more detailed information. In addition, we describe the modifications we have made to the model to allow for more flexible adjustments of its parameters.

\subsubsection{Review of the dust model} \label{size:review}

This model treats the galaxy as a one-zone object, in which dust enrichment is calculated in a consistent manner with metal enrichment. We apply it to individual subhalos.
The grain size distribution at time $t$ is denoted as $n(a,\, t)$ such that $n(a,\, t)\,\mathrm{d}a$ is the number density of dust grains per hydrogen atom in a grain radius range of $(a,\,a+\mathrm{d}a)$. We solve the evolution of $n(a,\, t)$ with an initial condition of
$n(a,\,t=0)=0$ (no dust). The total dust abundance is traced by the dust-to-gas ratio, which is evaluated by

\begin{align}
D=\int_0^\infty\frac{4}{3}\upi a^3sn(a)\,\mathrm{d}a/(\mu m_\mathrm{H}),
\end{align}

\noindent where $s$ is the material mass density of dust, $\mu =1.4$ is the gas mass per hydrogen atom, and $m_\mathrm{H}$ is the mass of the hydrogen atom.
Note that we do not distinguish between different grain species in the calculation of the grain size distribution because of the difficulty in treating the interaction among different species. Later, in the calculation of extinction curves, it will be necessary to distinguish multiple species (Section~\ref{model:ext}). For the calculation of grain size distribution, we adopt dust properties of graphite following HM20 (e.g.\ $s=2.24$ g cm$^{-3}$; \citealt{Weingartner:2001apj}). However, since the calculation of the grain size distribution is not sensitive to the assumed grain species, adopting a specific species does not affect the resulting size distributions significantly.

We consider the following five processes for dust evolution: stellar dust production (dust condensation in stellar ejecta), dust destruction in SN shocks in the ISM, dust growth by the accretion of gas-phase metals and by coagulation in the dense ISM, and dust disruption by shattering in the diffuse ISM. The evolution of the grain size distribution is calculated for each of the above processes as summarized below.

The stellar dust production is based on the metallicity evolution given by TNG. We estimate the increase of dust-to-gas ratio for a galaxy between two successive simulation snapshots from the increment of metallicity, by assuming the metal-to-dust condensation fraction, denoted as $f_\mathrm{in}$, to be 0.1. When the metallicity decreases because of e.g.\ infall of low-metallicity gas we manually dilute the dust in proportion to the decrease in the metallicity (further details are given in Section \ref{combined:accretion}). We assume that the newly-formed dust mass arising from stellar evolution has a log-normal grain mass distribution with an average grain radius of 0.1 $\micron$ and a standard deviation of 0.47 (in the log). This assumes that the dust grains produced by stars have submicron sizes (see \citealt{2019DustModel} and references therein for a discussion of arguments in support of this assumption). For dust destruction by SN shocks, we adopt a grain-size-dependent destruction efficiency, which is multiplied by the rate at which the interstellar dust is swept up by SN shocks.

The remaining three mechanisms only occur in specific phases of the ISM. Thus, we consider two ISM phases defined by the hydrogen number density $n_{\rm{H}}\, (\rm{cm^{-3}})$ and the gas temperature $T_{\rm{gas}}\, (\rm{K})$. One is the diffuse/warm ISM with $(n_{\rm{H}},\, T_{\rm{gas}}) \sim (10^{-1},\, 10^4)$, and the other is the dense/cold ISM with $(n_{\rm{H}},\, T_{\rm{gas}}) \sim (10^2,\, 10^2)$. In principle, the mass of gas in each phase could be determined from the density and temperature distributions of all the gas elements associated with each subhalo in the original simulation. However, we instead treat the fraction of gas in each phase as a parameter of our model and adopt the following fixed values for the density and temperature in each phase: $(n_{\rm{H}},\, T_{\rm{gas}}) = (0.3,\, 10^4)$ for the diffuse ISM and $(n_{\rm{H}},\, T_{\rm{gas}}) = (300,\, 100)$ for the dense ISM.

We apply this treatment for the following reasons. First, dense gas is not directly resolved in the simulation and instead relies on a sub-resolution two-phase model \citep{Springel03}. Thus, coagulation and accretion need to be treated with sub-grid prescriptions \citep[e.g.][]{Aoyama:2018mnras}. Second, the density and temperature are degenerate with the assumed turbulent velocities in coagulation and shattering, and uncertainty in the turbulent velocities can be absorbed in the coagulation and shattering efficiencies that are adjustable in this work. We will discuss more details in the next section. For these reasons, we simply adopt a fixed characteristic density and temperature for each ISM phase.

We denote the mass fraction of the dense ISM as $\eta_{\rm{dense}}$ and thus the mass fraction of the diffuse ISM is $1-\eta_{\rm{dense}}$. These fractions are included in the calculation in the form of an effective time-step. Let $\Delta t$ be the time-step in our calculations. We distribute this time-step to the dense and diffuse ISM by adopting $\eta_\mathrm{dense} \Delta t$ for the processes occurring in the dense ISM (accretion and coagulation) and $(1-\eta_\mathrm{dense})\Delta t$ for the process in the diffuse ISM (shattering). The evolution of the grain size distribution by accretion is solved by considering the grain growth rate dependent on the metallicity and the grain radius. The change of the grain size distribution by coagulation and shattering are treated by the Smoluchowski-type equations with kernel functions determined consistently with the grain--grain collision rates in a turbulent media. The turbulent velocity in each ISM phase is given by the model of \citet{Ormel:2009aa} albeit with an adjusted velocity normalization.

We discretize the entire grain radius range ($a=0.3$~nm--10~$\micron$) into 64 logarithmic bins. We apply $n(a,\, t)=0$ at the upper and lower radii as boundary conditions. That is, we remove grains that evolve beyond the above radius range. The interval between TNG snapshots is often longer than the characteristic time-scale of interstellar dust processing, in which case we divide the time interval into many sub-steps, each of which is shorter than the timescale of the most rapid process.

\subsubsection{Modification of the dust model: efficiencies} \label{size:modify}

Shattering and coagulation are both associated with grain-grain collision in the ISM. As mentioned above, the collision rate is regulated by the kernel function, which is a product of grain cross-section and grain velocity, in the Smoluchowski-type equations. HM20 simply used the geometrical cross-sections of compact spheres. The cross-sections depend on the grain shape. As mentioned above, the grain velocity also depends on the adopted turbulence model. Since the balance between coagulation and shattering determines the shape of grain size distribution in evolved (metal-enriched) galaxies, it is worth adjusting the coagulation and shattering efficiencies to clarify the robustness and uncertainty in the model.
To regulate the shattering and coagulation efficiencies, we introduce dimensionless parameters $\omega_\mathrm{shat}$ and $\omega_\mathrm{coag}$ and multiply the respective kernel functions. These parameters are useful to modify the relative strength between coagulation and shattering. Setting $\omega_\mathrm{shat} = \omega_\mathrm{coag} = 1$ gives the same rates as in HM20.

\subsection{Post-processing of the simulation data} \label{model:combined}

Two further significant modifications are required to this model, to treat subhalo growth through merging and smooth gas accretion, and to determine the dense gas fraction $\eta_\mathrm{dense}$. 

To handle mergers, we post-process the merger trees of each MW-like galaxy in our sample selected at $z=0$, following them back to the earliest recorded time of $z=12$. We first extract the necessary ISM properties of each progenitor subhalo. We then calculate the evolution of the grain size distribution using the model described in Section \ref{model:size}, tracing down the tree from the multiple leaf nodes to the single MW-like root node at $z=0$. In doing so, we compute the dust-to-gas ratio for each subhalo at each time. If a subhalo does not contain gas, it does not affect the dust abundance. Thus, subhalos without gas do not  influence our results.

\subsubsection{Dense gas fraction} \label{combined:dense_frac}

Determination of the dense gas fraction $\eta_\mathrm{dense}$ is critical since some processes occur only in one of the two ISM phases. Given the resolution limitations of cosmological simulations such as TNG, we do not measure this directly from the hydrodynamical outputs, but instead assume a relation between $\eta_\mathrm{dense}$ and the SFR:
\begin{equation} \label{eq:KSlaw}
    \eta_\mathrm{dense}=\tau_\mathrm{ff} \frac{\mathrm{SFR}}{\epsilon_* M_\mathrm{gas}^*},
\end{equation}
where $M_\mathrm{gas}^*$ is the interstellar gas mass, $\epsilon_*$ is the star-formation efficiency, and $\tau_\mathrm{ff} = 2.51 (n_\mathrm{H}/300 ~ \mathrm{cm}^{-3})^{-1/2}$ Myr is the free-fall time. We obtain $M_\mathrm{gas}^*$ and SFR for each subhalo from TNG300-1, and fix $\epsilon_*=0.01$ \citep{Krumholz2005_epsilon}. For  consistency with the assumed density of the dense ISM, we fix $n_\mathrm{H}=300$ cm$^{-3}$ in the estimation of the free-fall time.

As shown later, the resulting grain size distributions are sensitive to the choice of $\eta_\mathrm{dense}$. Thus, we also examine some models with fixed values of $\eta_\mathrm{dense}$ to understand the response of the results to the change of $\eta_\mathrm{dense}$ (Section \ref{results:parameter}).

\subsubsection{Dust evolution due to gas accretion} \label{combined:accretion}

Galaxies in TNG grow through hierarchical merging as well as via the smooth accretion of baryons and dark matter from the cosmic web. HM20 considered the dust evolution in an isolated field galaxy and did not account for either mode of growth. Thus, we need to extend HM20's model to deal with subhalo growth through smooth accretion (we hereafter refer to this process as gas accretion to avoid confusion with dust growth by accretion) and merging.

To model gas accretion, we note that galaxies accrete gas from the circumgalactic medium (CGM). We assume that the CGM contributes no dust and only few metals. Under this assumption, gas accretion from the CGM simply dilutes the dust-to-gas ratio. We therefore regard decreasing metallicity as an indicator of significant gas accretion. 
Furthermore, we assume that neither the dust-to-metal ratio nor the functional shape of the grain size distribution changes during a gas accretion episode. Thus, if the gas accretion changes the metallicity from $Z(t)$ to $Z(t+\Delta t)[<Z(t)]$, we update the grain size distribution $n(a,\, t)$ according to the ratio between $Z(t)$ and $Z(t+\Delta t)$. That is, the grain size distribution at $t+\Delta t$ is evaluated as
\begin{equation} \label{eq:dilution}
    n(a,\,t+\Delta t) = n(a,\,t)\,\frac{Z(t+\Delta t)}{Z(t)}~~~\text{(in a gas accretion episode)}.
\end{equation}
Note that we only evolve the dust properties in this way when no mergers occur in a given time-step.

\subsubsection{Dust evolution due to mergers} \label{combined:merger}

\begin{figure}
  \centering
  \includegraphics[width=\columnwidth]{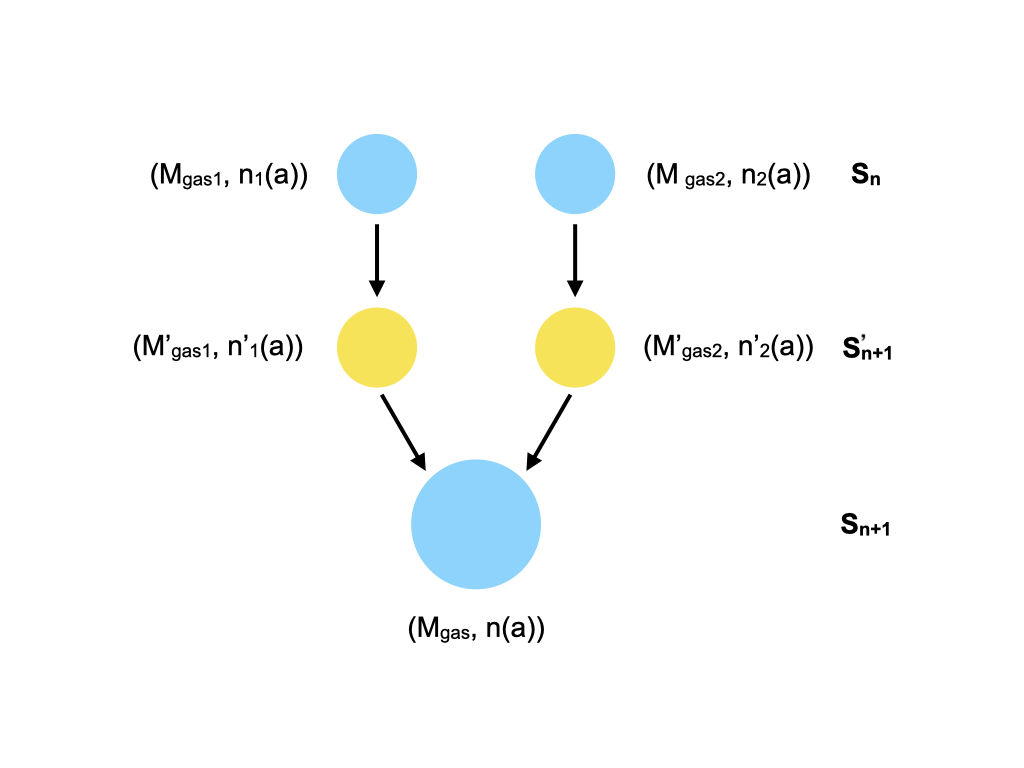}
  \caption{Illustration of our treatment for grain size distribution evolution in mergers. The blue circles show TNG subhalos while the yellow circles are subhalos inserted virtually for the purpose of the merging calculation. $S_n$ denotes the $n$th snapshot in TNG and ($M_{\rm{gas}},\,n(a)$) are the pair of gas mass and grain size distribution for a given subhalo at one snapshot.} \label{fig:merger}
\end{figure}

Although TNG provides information of progenitors and descendants for each subhalo, we do not know exactly when a merger happens because TNG stores snapshot data with a time interval of roughly 0.1 Gyr. To overcome this limitation we assume that (i) the merger happens at the end of the interval between two snapshots, (ii) the amount of accreting gas mass right before the merger is proportional to the gas mass in the progenitor, and (iii) the total gas mass is conserved during the merging. In practice, we insert a virtual node right before the descendant subhalo, evolve grain size distributions from the progenitor to this virtual node, and combine the grain size distributions from two virtual nodes. 

Our method is illustrated in Fig.~\ref{fig:merger}. $S_n$ denotes the $n$th snapshot, with $n$ increasing with cosmic time. Blue circles are the original subhalos from TNG and yellow circles show the virtual nodes. The unprimed quantities, $M_\mathrm{gas}$ and $n(a)$, stand for the total gas mass directly obtained from TNG and the grain size distribution calculated by our model, respectively. We use the primed quantities $M_\mathrm{gas}'$ and $n'(a)$ for virtual nodes (pre-merging stage), and estimate them as follows. The progenitors (blue circles) in $S_{n}$ eventually evolve into the pre-merging stage (yellow circle) in $S_{n+1}'$. Under assumption (ii) we have $M_\mathrm{gas1}'/M_\mathrm{gas2}' = M_\mathrm{gas1}/M_\mathrm{gas2}$. Additionally, under assumption (iii), we have $M_\mathrm{gas1}' + M_\mathrm{gas2}' = M_\mathrm{gas}$. Hence, the grain size distribution in the merged system $n(a)$ is estimated by the following average:
\begin{equation} \label{eq:dust_merge}
    n(a) = \frac{M_\mathrm{gas1}'n'_1(a) + M_\mathrm{gas2}'n'_2(a)}{M_\mathrm{gas}}.
\end{equation}
The evolution from $n_i(a)$ to $n_i'(a)$ is calculated using our dust evolution model, as described in Section \ref{model:size}. In this way the information available at  snapshot $S_n$ allows us to obtain the grain size distribution at $S_{n+1}$.


\subsection{Extinction curve} \label{model:ext}

The above grain size distributions are calculated for the total dust abundance. In order to calculate the observational properties of dust, it is important to specify the grain species. In this paper, we calculate the extinction curve as a representative observable quantity that reflects the grain size distribution. The extinction per hydrogen at wavelength $\lambda$, $A_\lambda$, is calculated as 
\begin{equation} \label{eq:extinction}
    A_\lambda = 2.5 \log_{10} \mathrm{e} \sum_i \int^\infty_0 n_i(a) \pi a^2 Q_\mathrm{ext}^{(i)}(a,\lambda)\;da,
\end{equation}
where \textit{i} indicates the different grain species, $Q_\mathrm{ext}^{(i)}(a,\lambda)$ is the extinction efficiency factor of the $i$th species evaluated with Mie theory \citep{Bohren:1983book}. For the grain species, following HM20, we consider silicate, aromatic carbon, and non-aromatic (amorphous) carbon. We adopt the astronomical silicate taken from \citet{Weingartner:2001apj} (originally \citealt{DraineLee1984apj, LaorDraine1993apj}), while we apply graphite in the same paper for the aromatic component. For the non-aromatic component, we adopt the optical constants of amorphous carbon given by `ACAR' in \citet{Zubko1996mnrasAC}. Other sets of grain materials could be used \citep[e.g.][]{Zubko:2004apjs,Jones:2017aa}, but since the MW extinction curve is reproduced quite well in HM20, we adopt the same dust properties to concentrate on the differences caused by the inclusion of hierarchical structure formation.

As the grain size distribution calculated in our model is the total of all the dust species we must decompose it into these components. We assume the size distribution of each component to have the same shape but different abundance. Therefore, $n_i(a)$ differs from component to component only by a factor of the relative abundance, determined as follows. The fraction of silicate grains is calculated from the mass ratio of Si and C,
\begin{equation} \label{eq:metal_ratio}
    f_\mathrm{Si} = \frac{6Z_\mathrm{Si}}{6Z_\mathrm{Si} + Z_\mathrm{C}}.
\end{equation}
This implicitly assumes that the dust couples with the gas and has similar chemical composition. The coefficient 6 is due to the mass fraction of Si in silicate (1/6) \citep{Hirashita:2011mnras}. We define the aromatic fraction $f_\mathrm{ar}$ as the fraction of aromatic carbon to the total carbon dust. HM20 showed, based on \citet{Murga:2019mnras}, that $f_\mathrm{ar}$ quickly reaches an equilibrium whereby $f_\mathrm{ar} = 1 - \eta_\mathrm{dense}$ for most grain radii \citep[see also][]{Rau:2019mnras}. Hence, in this work we fix the aromatic fraction as $f_\mathrm{ar} = 1 - \eta_\mathrm{dense}$.


\section{Results}\label{sec:results}

Since our dust evolution model contains several parameters, we first examine the parameter dependence in Section~\ref{results:parameter}. For the survey of parameter values, we focus on a subset of the full MW-like galaxy sample. Based on this survey we determine a final parameter set, which is then applied to the full sample. The full statistical results are presented in Section \ref{results:stat}.

\subsection{Parameter dependence}\label{results:parameter}

We test the parameter dependence of key results for a small subset of the full MW-like sample. The subset contains ten galaxies and is randomly selected from the full sample. To illustrate the evolution of dust properties with redshift in each sample subhalo, we only show the grain size distributions along the main branch.

\subsubsection{Gas mass associated with the dust evolution} \label{param:gas_mass}

As explained in Section~\ref{model:TNG}, we consider two definitions of gas mass associated with the dust evolution: one is the the total gas mass within the subhalo ($M_\mathrm{gas,tot}$), and the other is the gas mass within twice the stellar half mass radius ($M_\mathrm{gas,2R*}$). We compare these two alternatives to investigate the uncertainty caused by the definition of gas mass associated with star formation and dust processing in galaxies.

\begin{figure}
    \centering
    \includegraphics[width=\columnwidth]{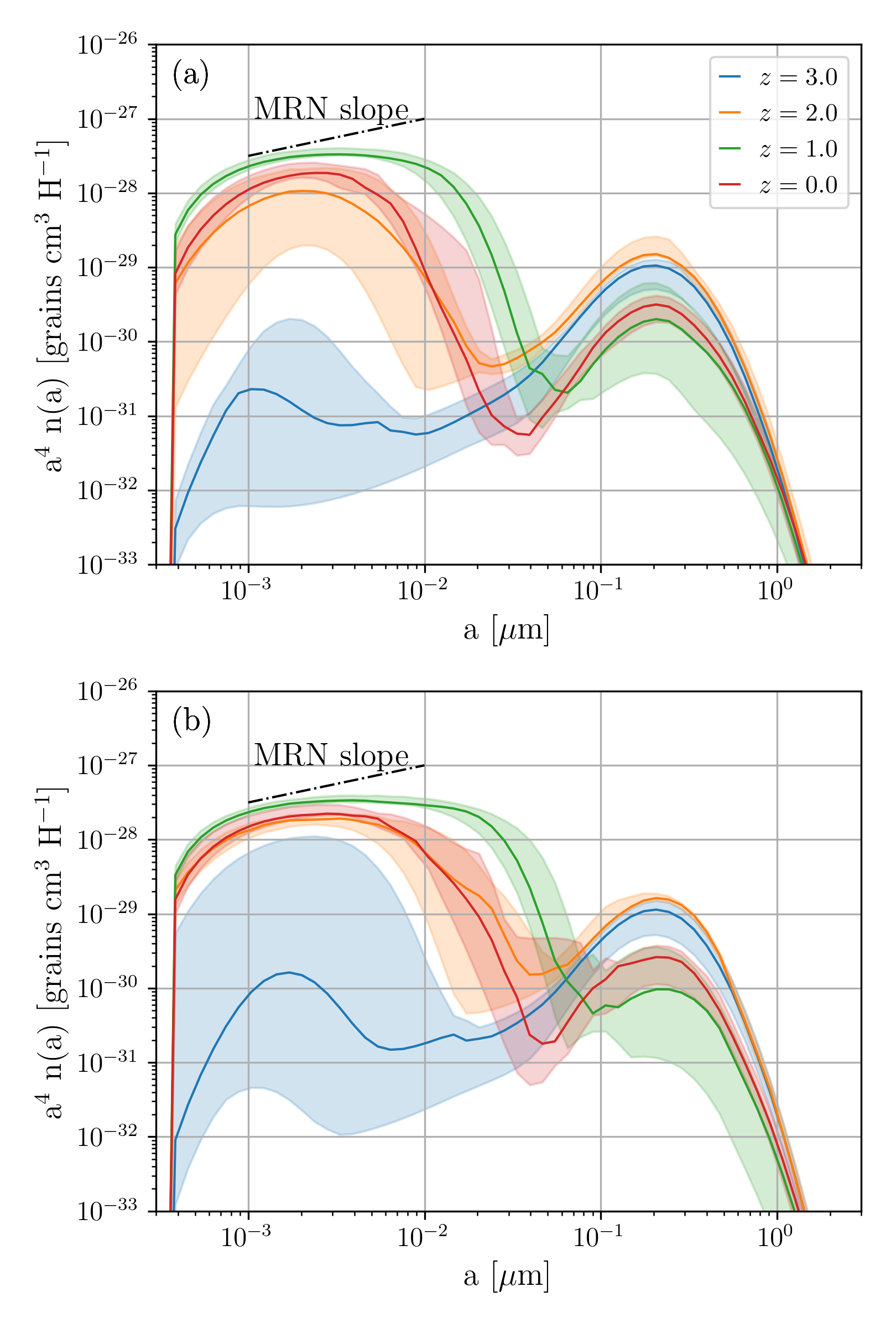}
    \caption{Grain size distributions in the main branch of a small subsample of galaxies with different gas masses taken for the dust model. Panels (a) and (b) show the results for using the total gas mass within the subhalo, $M_\mathrm{gas,tot}$, and the gas mass within twice the stellar half mass radius, $M_\mathrm{gas,2R*}$, respectively. The solid curve is the median and the shaded area is the value between 25th and 75th percentiles in each size bin. The results at $z=3$, 2, 1, and 0 are shown in blue, orange, green and red, respectively. The dash--dotted black line shows the MRN slope for a reference. The grain size distribution per hydrogen atom, $n(a)$, is multiplied by $a^4$ so that the resulting quantity is proportional to the dust abundance per $\log a$.}
    \label{fig:gas_mass}
\end{figure}

In Fig.\ \ref{fig:gas_mass} we show the grain size distributions obtained using $M_\mathrm{gas,tot}$ and $M_\mathrm{gas,2R*}$ in Panels (a) and (b), respectively. The case with $M_\mathrm{gas,tot}$ tends to have more intermediate-sized ($a\sim 0.03~\micron$) grains than the case with $M_\mathrm{gas,2R*}$ at $z\lesssim 1$. Although the grain size distribution has a large dispersion for small grains at early epochs, the grain size distributions are not significantly affected by the choice of gas mass for large grains or at later epochs. The sensitive behaviour at early epochs is due to dust growth by accretion, which causes a drastic increase of small grains on a short time-scale. However, the overall median behaviour is not strongly affected by the choice of gas mass, which motivates our fiducial choice of $M_\mathrm{gas,2R*}$ for our modeling below.

\subsubsection{Dense gas fraction} \label{param:dense_gas}

\begin{figure}
    \centering
    \includegraphics[width=\columnwidth]{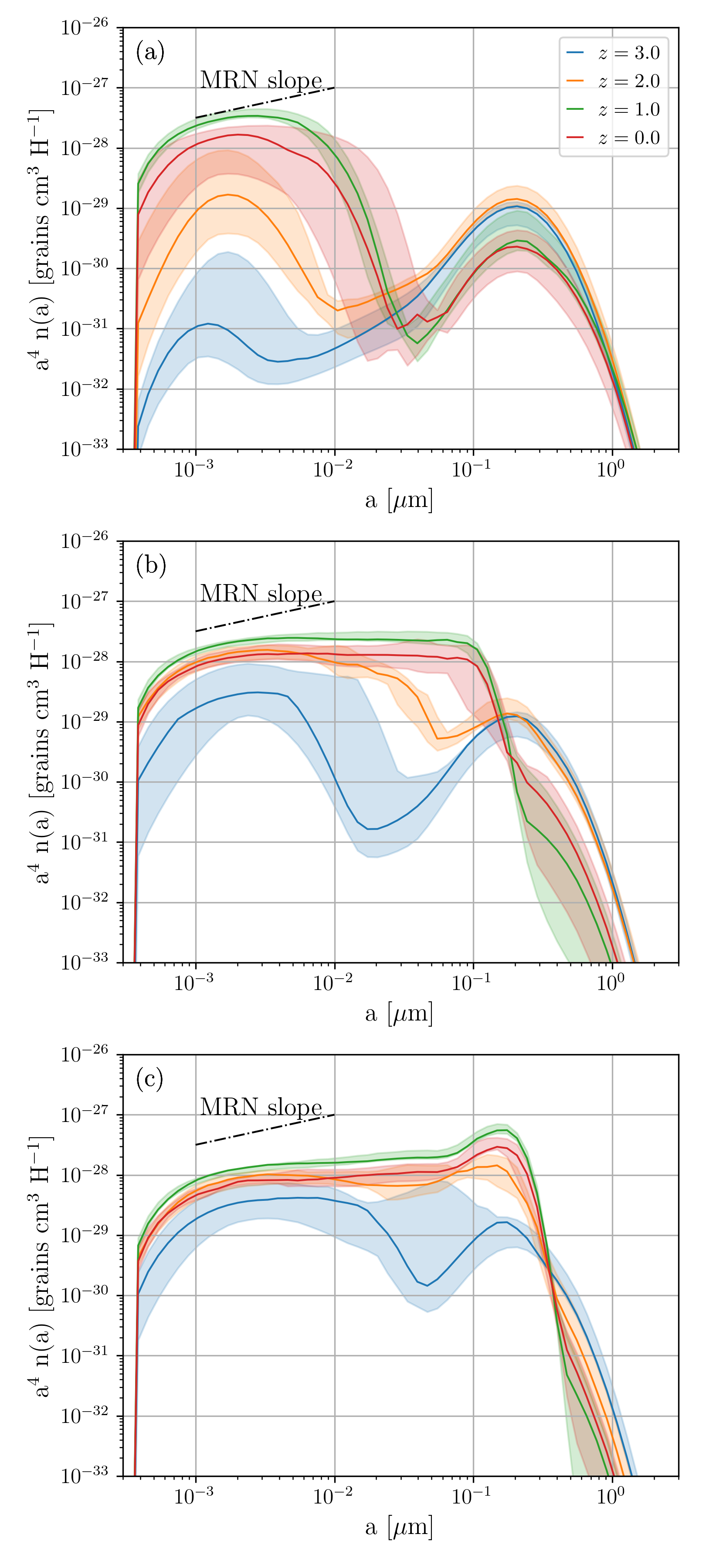}
    \caption{Same as Fig.\ \ref{fig:gas_mass} but for various values of $\eta_\mathrm{dense}$, which is assumed to be constant here: (a) $\eta_\mathrm{dense} = 0.1$, (b) $\eta_\mathrm{dense} = 0.3$, and (c) $\eta_\mathrm{dense} = 0.5$.}  \label{fig:fixed_frac}
\end{figure}

The dense gas fraction $\eta_\mathrm{dense}$ affects dust evolution processes, especially accretion and coagulation, which are more significant in the dense ISM. To demonstrate the impact of dense gas fraction, we first fix the dense gas fraction at different values ($\eta_\mathrm{dense}=0.1$, 0.3, and 0.5) and show the results in Fig.\ \ref{fig:fixed_frac}. It is clear that different dense gas fractions not only change the dust abundance but also alter the grain size distribution. A larger value of $\eta_\mathrm{dense}$ leads to an enhanced small-grain abundance in the early phases because accretion is more efficient. At later epochs, we observe a higher abundance of large grains for higher $\eta_\mathrm{dense}$ because coagulation is more efficient. With $\eta_\mathrm{dense} = 0.5$, the overall slope of grain size distribution approaches the MRN distribution, which is consistent with the results of one-zone models in HM20. 

\subsubsection{Shattering and coagulation} \label{param:shat_coag}

\begin{figure}
    \centering
    \includegraphics[width=\columnwidth]{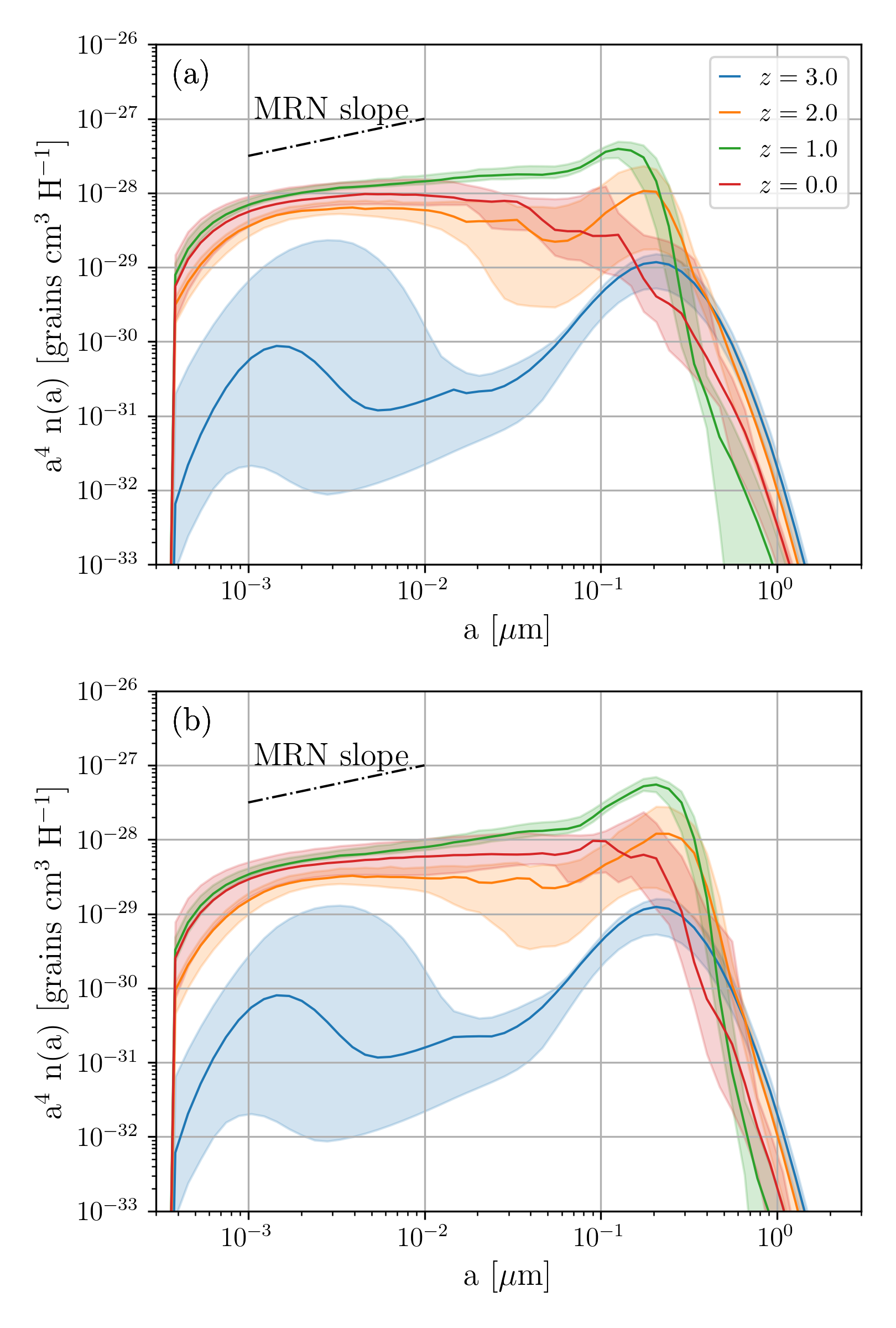}
    \caption{Same as Fig.\ \ref{fig:gas_mass} but with different coagulation efficiencies: (a) $\omega_\mathrm{coag}=5$ (b) $\omega_\mathrm{coag}=10$.} \label{fig:coag_results}
\end{figure}

\begin{figure}
    \centering
    \includegraphics[width=\columnwidth]{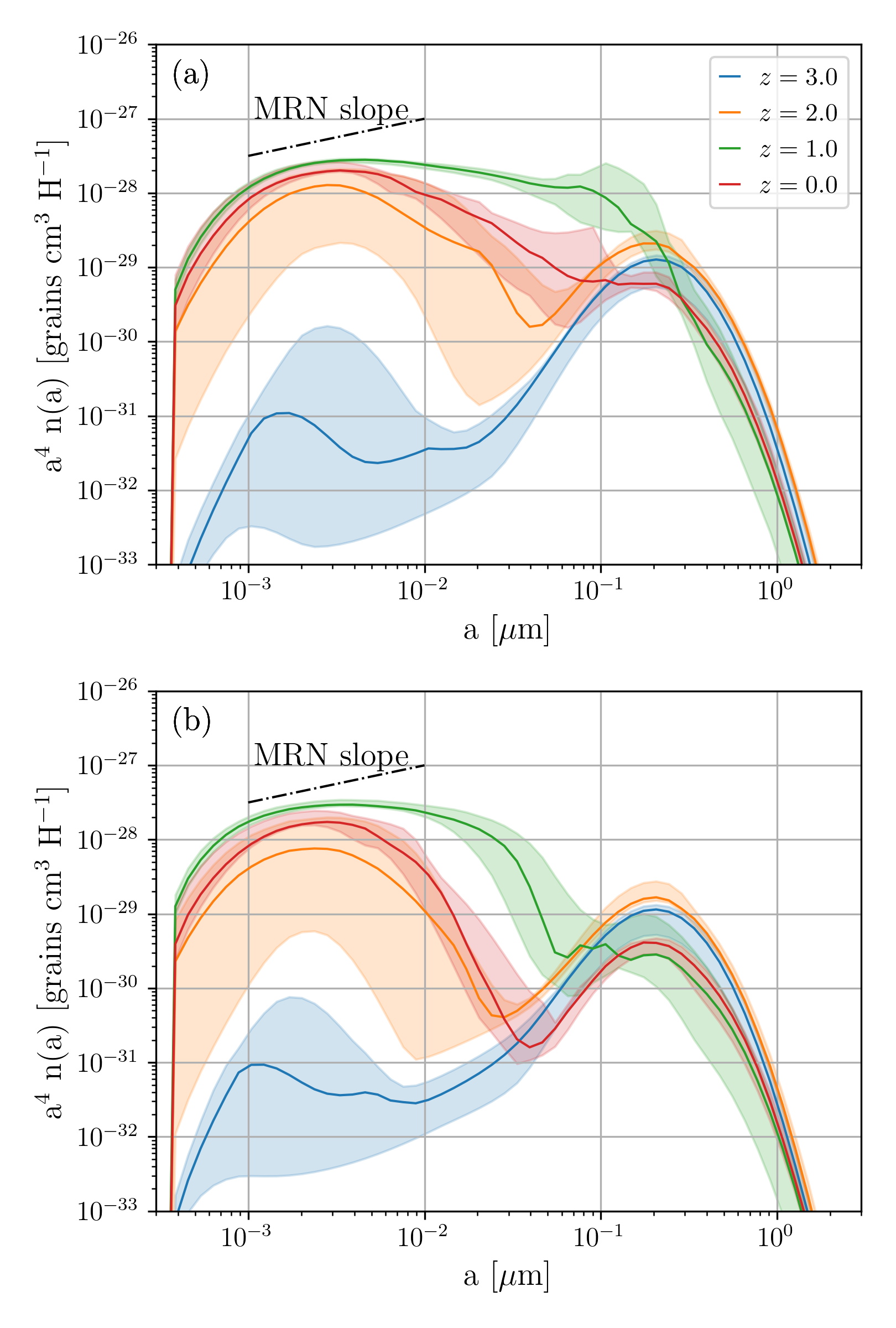}
    \caption{Same as Fig.\ \ref{fig:gas_mass} but with different shattering efficiencies: (a) $\omega_\mathrm{shat}=0.2$ and (b) $\omega_\mathrm{shat}=0.5$.} \label{fig:shat_results}
\end{figure}

The shape of the grain size distribution is affected by the efficiency of shattering as well as coagulation, especially at later stages. As introduced in Section \ref{size:modify}, we adjust the strength of coagulation ($\omega_\mathrm{coag}$) and shattering ($\omega_\mathrm{shat}$) to investigate if we obtain grain size distributions more similar to the MRN distribution at $z=0$. We show the results in Figs.\ \ref{fig:coag_results} and \ref{fig:shat_results} for coagulation and shattering, respectively. We find that the enhancement of the coagulation efficiency increases the large grain abundance and tends to produce a MRN-like slope at $z\lesssim 1$. On the other hand, although a reduction of shattering strength increases the large-grain abundance, this is still not enough to reproduce the MRN slope
(Fig.\ \ref{fig:shat_results}b). Efficient coagulation is essential to produce a slope similar to the MRN distribution as noted in \citet{2019DustModel}. Note that neither $\omega_\mathrm{coag}$ nor $\omega_\mathrm{shat}$ affects the grain size distributions at $z\gtrsim 3$ because these processes only play a minor role at that early epoch.

In light of the results in Figs.\ \ref{fig:coag_results} and \ref{fig:shat_results}, the enhancement of coagulation might be the more appropriate way to reproduce the MRN slope. Although our purpose is not to reproduce the MRN grain size distribution exactly, it is still worth exploring which solutions yield grain size distributions similar to that of the Milky Way. Considering that grain cross-sections and grain velocities are uncertain to an order of magnitude, we use $\omega_\mathrm{coag}=10$ as the fiducial model for the full sample. To minimize fine-tuning we fix $\omega_\mathrm{shat}=1$.
Despite this tuning, we still expect that the physical interpretations given below will hold for any simulations in the future, at least qualitatively. Since there have been no calculations of full grain size distributions in cosmological simulations, our fiducial model serves as a robust basis for investigating which process dominate the evolution of the grain size distribution in the MW-like galaxies at various epochs.

\subsubsection{Choice of the fiducial parameters} \label{param:fiducial}

Based on the above tests, we fix parameters for the full-sample calculations described in the next subsection. As shown in Section~\ref{param:gas_mass}, there are two definitions of galactic gas mass available in TNG, but the choice between these does not significantly affect the grain size distributions predicted by our model. Since most of the dust formation and destruction processes occur around stars, rather than in the dark matter halo, we use the gas mass within twice the stellar half-mass radius in our dust model. Hereafter, we simply denote $M_\mathrm{gas,2R}$ as $M_\mathrm{gas}$.

We assumed various constant values for the dense-gas fraction ($\eta_\mathrm{dense}$) in Section \ref{param:dense_gas} for the purpose of testing the effect of this parameter. However, we do not fix $\eta_\mathrm{dense}$ in our fiducial model, but instead use the recipe described in Section \ref{combined:dense_frac} (equation \ref{eq:KSlaw}) to determine its value in each subhalo individually. As discussed in Section \ref{param:shat_coag}, we adopt $\omega_\mathrm{coag}=10$, but leave the shattering efficiency unchanged ($\omega_\mathrm{shat}=1$).

\subsection{Statistical results for the full sample}\label{results:stat}

We proceed to model the full sample of 210 MW-like galaxies in TNG300-1. For four of these galaxies, we encounter non-physical values for the dust-to-gas ratio. We find that these galaxies have once merged with some subhalos of non-cosmological origins, which are probably the results of subhalo mis-classification. Thus, we discard these four merger trees and analyze the remaining 206, which we refer to as the full sample. Excluding those four galaxies does not affect our results and conclusions.

\subsubsection{Dust-to-gas ratio and metallicity} \label{stat:DZ}

\begin{figure}
    \centering
    \includegraphics[width=\columnwidth]{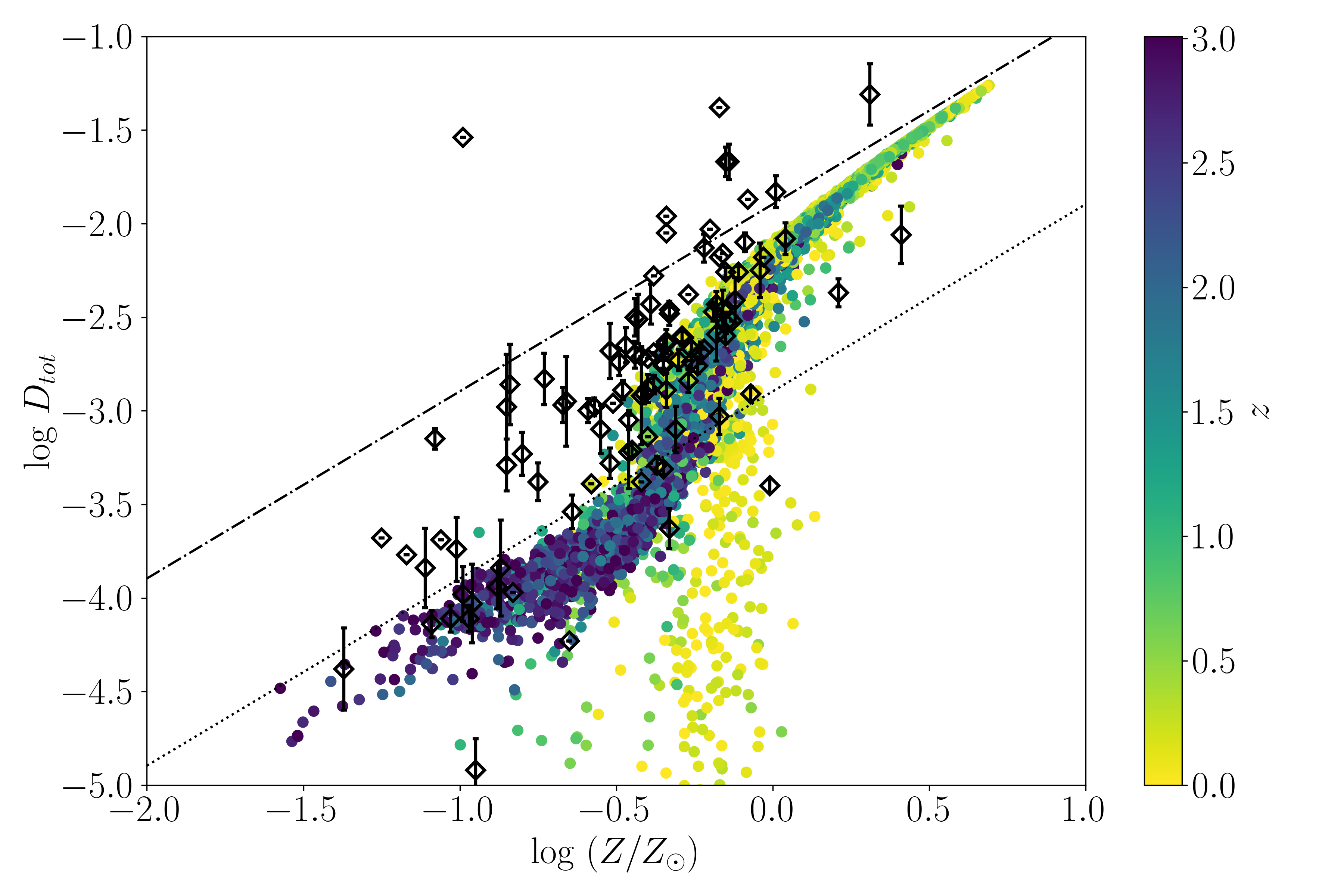}
    \caption{Relation between the dust-to-gas ratio and metallicity for our full galaxy sample and their progenitors from $z=0$ to 3 (colour indicating redshift). The gray open diamonds with error bars are observational data from the local Universe assuming a metallicity-dependent CO-to-H$_2$ conversion factor \citep[taken from][]{DZ_Remy2014}. The dash--dotted line denotes the saturation limit of the dust-to-gas ratio ($D_\mathrm{tot}=Z$), and the dotted line denotes the linear relation of the stellar yield ($D_\mathrm{tot}=0.1\,Z$).} \label{fig:dz_diagram}
\end{figure}

The relation between dust-to-gas ratio and metallicity has often been used to test dust enrichment models \citep[e.g.][]{Lisenfeld:1998aa,Dwek:1998aa}. To examine if our model reproduces the dust enrichment reasonably, Fig.\ \ref{fig:dz_diagram} shows the evolution of dust-to-gas ratio for all galaxies in our sample, together with their progenitors at earlier times, over the redshift range $z=0$--3. We compare the calculated $D_\mathrm{tot}$--$Z$ relation to observational results for local galaxies \citep{DZ_Remy2014}, for which the dust mass is estimated from IR-to-submm photometry, and the gas mass is estimated from H \textsc{i} and CO data with a metallicity-dependent CO-to-H$_2$ conversion factor.

The dust-to-gas ratios our model predicts at early times are slightly below the linear relation expected from the dust condensation efficiency in stellar ejecta ($=0.1$; dotted line in Fig.\ \ref{fig:dz_diagram}). Dust destruction by SNe is responsible for the shift towards lower dust-to-gas ratios. The dust-to-gas ratio and metallicity of galaxies increase with decreasing redshift, but the relation stays on a well defined narrow track, except for a stripe of low dust-to-gas ratios around $Z\sim 0.2$~Z$_{\sun}$. This relation between dust-to-gas ratio and metallicity is broadly consistent with observations, including the nonlinear (steep) increase of dust-to-gas ratio around $Z\sim 0.3$~Z$_{\sun}$ \citep[see also][for a recent analysis]{Aniano2020}. The fraction of points in the band toward low $D_\mathrm{tot}$ around $Z\sim 0.2$ Z$_{\sun}$ is only about 1 percent of all the points. We will discuss the reason for these outliers in Section \ref{sec:discussion}.

\begin{figure*}
    \centering
    \includegraphics[width=0.9\textwidth]{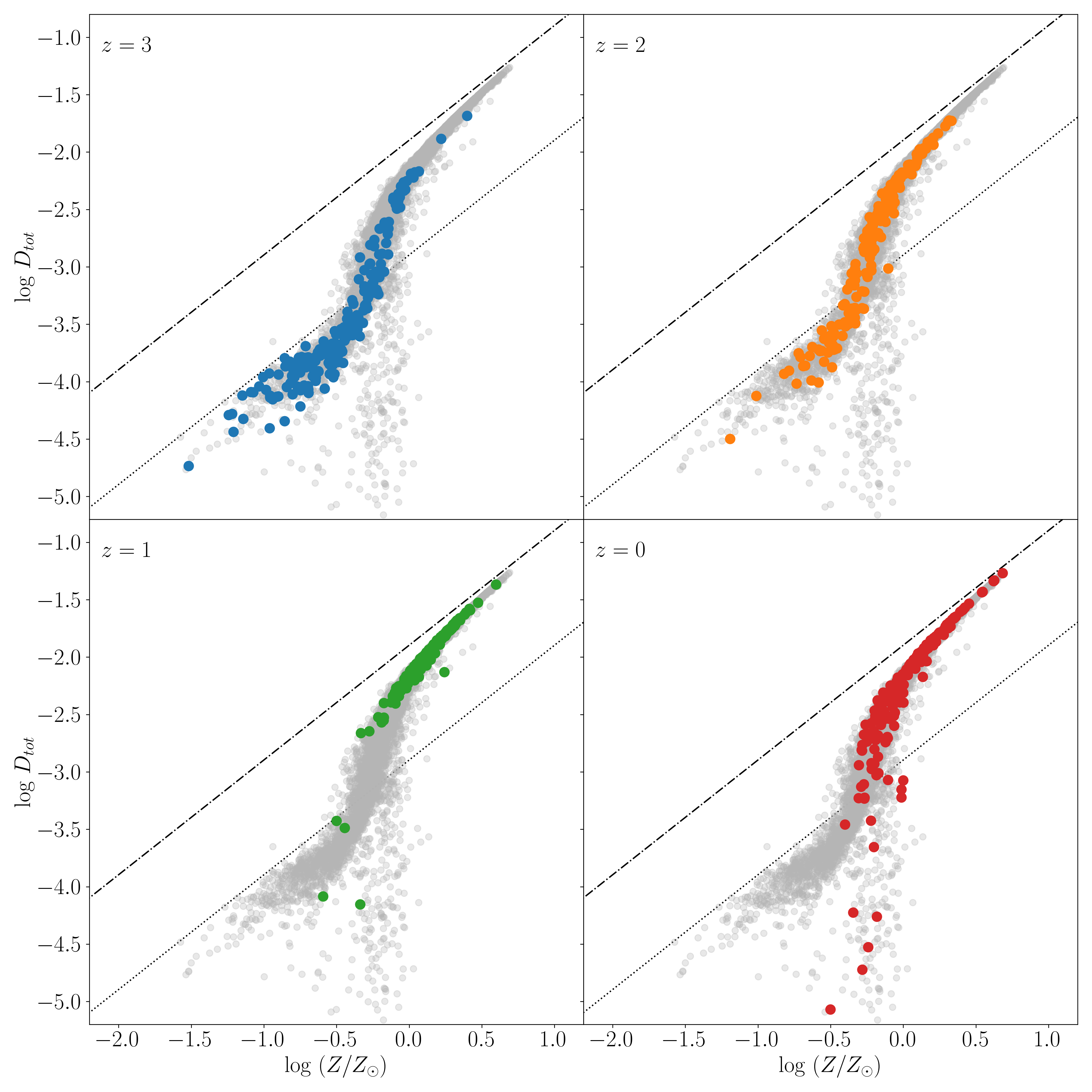}
    \caption{Dust-to-gas ratio versus metallicity for the galaxies in our sample, together with their progenitors at earlier times, shown at (a) $z=3$, (b) $z=2$, (c) $z=1$, and (d) $z=0$, each shown by colored circles. The gray background distribution shows all galaxies across all redshifts, as in Fig.\ \ref{fig:dz_diagram}. The dash-dotted line denotes the saturation limit of the dust-to-gas ratio ($D_\mathrm{tot}=Z$) and the dotted line shows the linear relation of the stellar yield ($D_\mathrm{tot}=0.1\,Z$).}
    \label{fig:separateDZ}
\end{figure*}

To illustrate the redshift evolution of the $D_\mathrm{tot}$--$Z$ relation more clearly, we plot the $D_\mathrm{tot}$--$Z$ diagram at $z=0$, 1, 2, and 3 separately in Fig.\ \ref{fig:separateDZ} (colored circles) overlaid on the combined distribution across all redshifts (gray background cloud). At $z=3$, most galaxies are near the line predicted from the stellar yield ($Z\sim f_\mathrm{in}D_\mathrm{tot}$), but some systems at $Z\gtrsim 0.3$ Z$_{\sun}$ have started to increase their dust abundance by accretion.
At $z=2$ the distribution moves upwards and to the right, and most galaxies are located in the regime where the increase in dust-to-gas ratio becomes non-linear. During this epoch, dust production is no longer dominated by stellar sources but instead by dust growth through accretion.

After this epoch, the distribution gradually approaches the upper-right corner of the diagram, near the saturation limit of $D_\mathrm{tot}=Z$ at $z=1$. It is clear from Fig.\ \ref{fig:separateDZ}c that the outliers start to appear at this epoch. Progressively more outliers appear at lower redshift, such that at $z=0$ the fraction reaches roughly 4 percent. Meanwhile, the tight correlation gradually broadens with larger scatter at $z=0$. As we will show later, the average metallicity of our sample reaches its peak at $z\sim0.5$ and then drops (Fig.\ \ref{fig:Z_evolution}a). This decrease of metallicity affects the dust-to-gas ratio, because we disable accretion when the gas-phase metallicity is decreasing (although SN destruction still takes place). The large dispersion of $\eta_\mathrm{dense}$ at $z\sim0$ (as shown later in Fig.\ \ref{fig:Z_evolution}b) could also increase the scatter in Fig.\ \ref{fig:separateDZ} because the dense gas fraction also affects the accretion and the relative strength of coagulation and shattering, which will have a mild effect on the total dust abundance.

In summary, our model is largely consistent with the observed relation between dust-to-gas ratio and metallicity. This indicates that we successfully reproduce the dust enrichment history of the MW-like galaxy sample. We further predict that the relation between dust-to-gas ratio and metallicity does not depend strongly on redshift. There is an increasing trend of dust-to-gas ratio with decreasing redshift, but the evolutionary track on the $D_\mathrm{tot}$--$Z$ is almost invariant. This is consistent with other theoretical models of dust enrichment in cosmological simulations by \citet{Hou:2019aa} (at $z\lesssim 2$) and \citet{Li2019}, and the semi-analytic model of \citet{Popping:2017aa} and \citet{Triani2020} (although \citealt{Vijayan2019} showed an additional dependence on the stellar age). The tight $D_\mathrm{tot}$--$Z$ relation is also compatible with the analytical result that metallicity is the main driver of dust enrichment \citep{Inoue:2011eps}.

\subsubsection{Grain size distributions}\label{stat:GSD}

\begin{figure}
    \centering
    \includegraphics[width=\columnwidth]{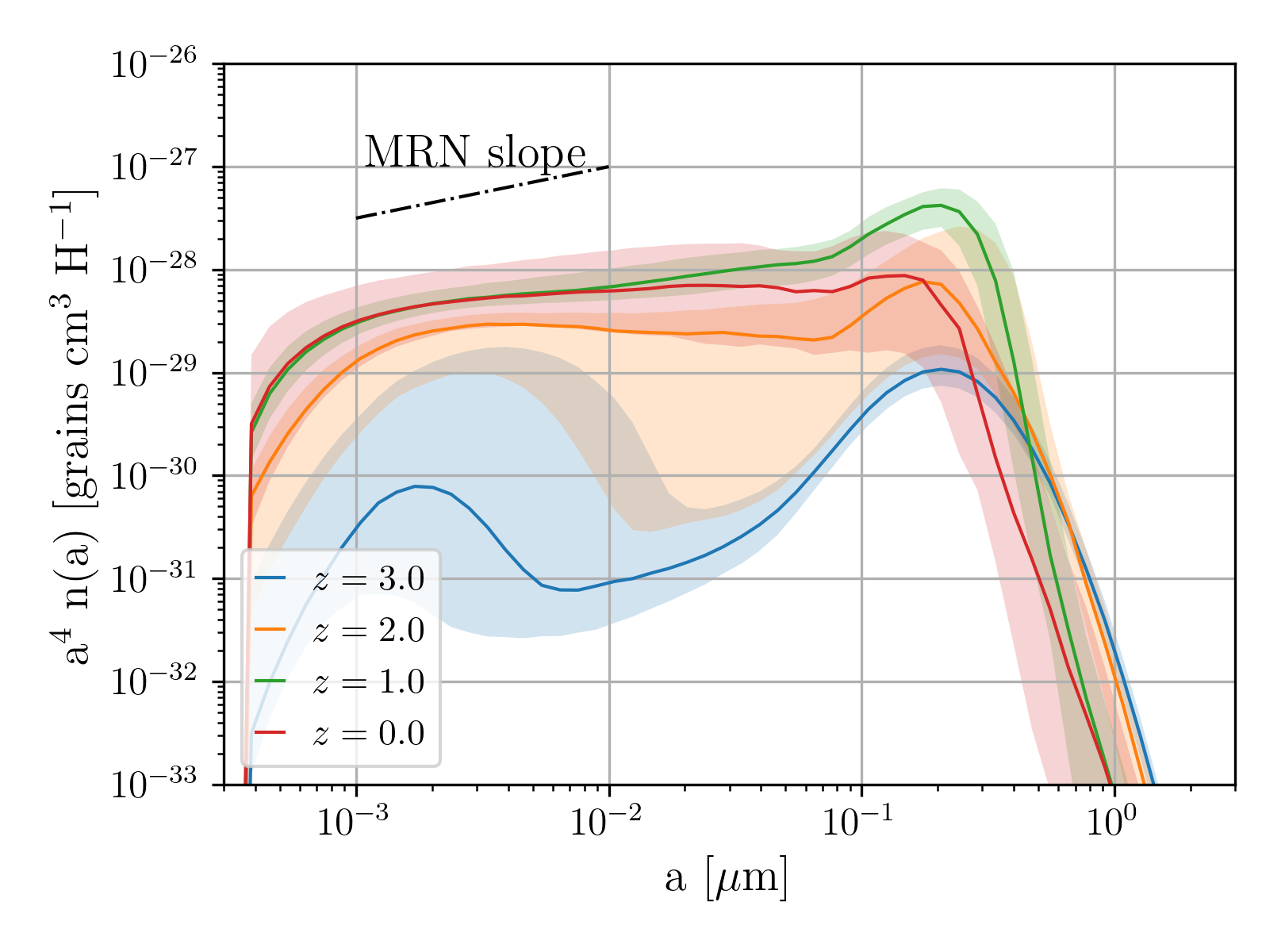}
    \caption{Grain size distributions of all 206 MW-like galaxies in our sample and their progenitors at different epochs. The blue, yellow, green, and red lines show the median in each grain radius bin at $z=3$, 2, 1, and 0, respectively. Shaded areas with the same colors show the corresponding 25th to 75th percentile ranges. The dot--dashed line shows the slope of the MRN grain size distribution $n(a)\propto a^{-3.5}$.} \label{fig:size_dist}
\end{figure}

We present the grain size distributions at different redshifts ($z=0$--$3$) for the progenitors of our MW-like galaxy sample in Fig.~\ref{fig:size_dist}. At $z>3$, the size distributions are dominated by stellar dust production, and their shapes are roughly described by a log-normal function centred at $a\sim 0.1~\micron$, as assumed for the grain size distribution of stellar dust (Section \ref{size:review}). To quantify the distribution we show the median with 25th and 75th percentiles at each grain radius at each redshift. We first discuss the evolving median behavior and then focus on the scatter in the sample. 

At $z=3$, the median distribution has two peaks, and the larger-grain maximum ($a \sim  0.2~\micron$)\footnote{The centre of the lognormal distribution is at $a=0.1~\micron$, but since we multiply the distribution by $a^4$, the peak of the distribution in the figure is located around $a\sim 0.2~\micron$.}
is higher than the peak in the small-size regime. The larger-grain peak is created by stellar dust production (Section \ref{size:review}), while the smaller-grain peak arises from dust growth by accretion, which is more efficient for small grains \citep{Kuo:2012mnras,2019DustModel}. The grain size distribution increases from $z=3$ to $z=2$ at almost all grain radii, especially for small sizes, reaching a nearly flat distribution at $a\lesssim 0.1~\micron$.

From $z=2$ to $z=1$ a further increase in dust abundance makes the integrated size distribution the largest among the four epochs. At $z=1$, because of efficient coagulation, the grain size distribution tends to a smooth power-law-like shape, approaching an MRN slope ($n\propto a^{-3.5}$). From $z=1$ to $z=0$ grains with $a\gtrsim 0.1~\micron$ are removed by shattering while the small-size end of the distribution remains nearly constant. As a result the grain size distributions we predict are more dominated by small grains than in the MRN distribution. Thus, shattering plays an important role in the grain size distribution at $z\sim 0$, as we discuss further below.

The scatter of the grain size distribution, shown by 25th--75th percentiles in Fig.\ \ref{fig:size_dist}, can arise from the different ISM properties in each galaxy. At $z=3$, the scatter in the grain size distribution is small at large radii, where the grain abundance is dominated by stellar dust production. The dispersion at large grain sizes is roughly the same as in the metallicity at $z=3$. On the other hand, the dispersion at small grain radii is large. Since the peak in the grain size distribution at small radii is caused by accretion (dust growth), the large scatter reflects variation in the fraction of dense gas in which that process occurs.

At $z=2$, the scatter becomes larger but then decreases by $z=1$. Thus, the typical epoch at which the small-grain abundance increases is around $z\sim 2$, and this is predominantly driven by accretion, which also induces a scatter at large grain radii through coagulation. The grain size distributions converge to a smooth MRN-like shape at $z\sim 1$, with extremely small scatter. This indicates the universality of the grain size distribution achieved as a result of efficient interstellar processing. The scatter tends to be larger at large grain radii if coagulation is weaker (Section \ref{results:parameter}). This means that coagulation plays an important role in realizing this `universality' of the grain size distribution.
At $z=0$, the scatter becomes larger again, especially at large grain radii. This is driven by the decrease of large grains by shattering and inefficient coagulation, due to a decrease in the dense gas fractions of the simulated galaxies, as discussed further below.

\subsubsection{Extinction curves} \label{stat:ext}

\begin{figure}
    \centering
    \includegraphics[width=\columnwidth]{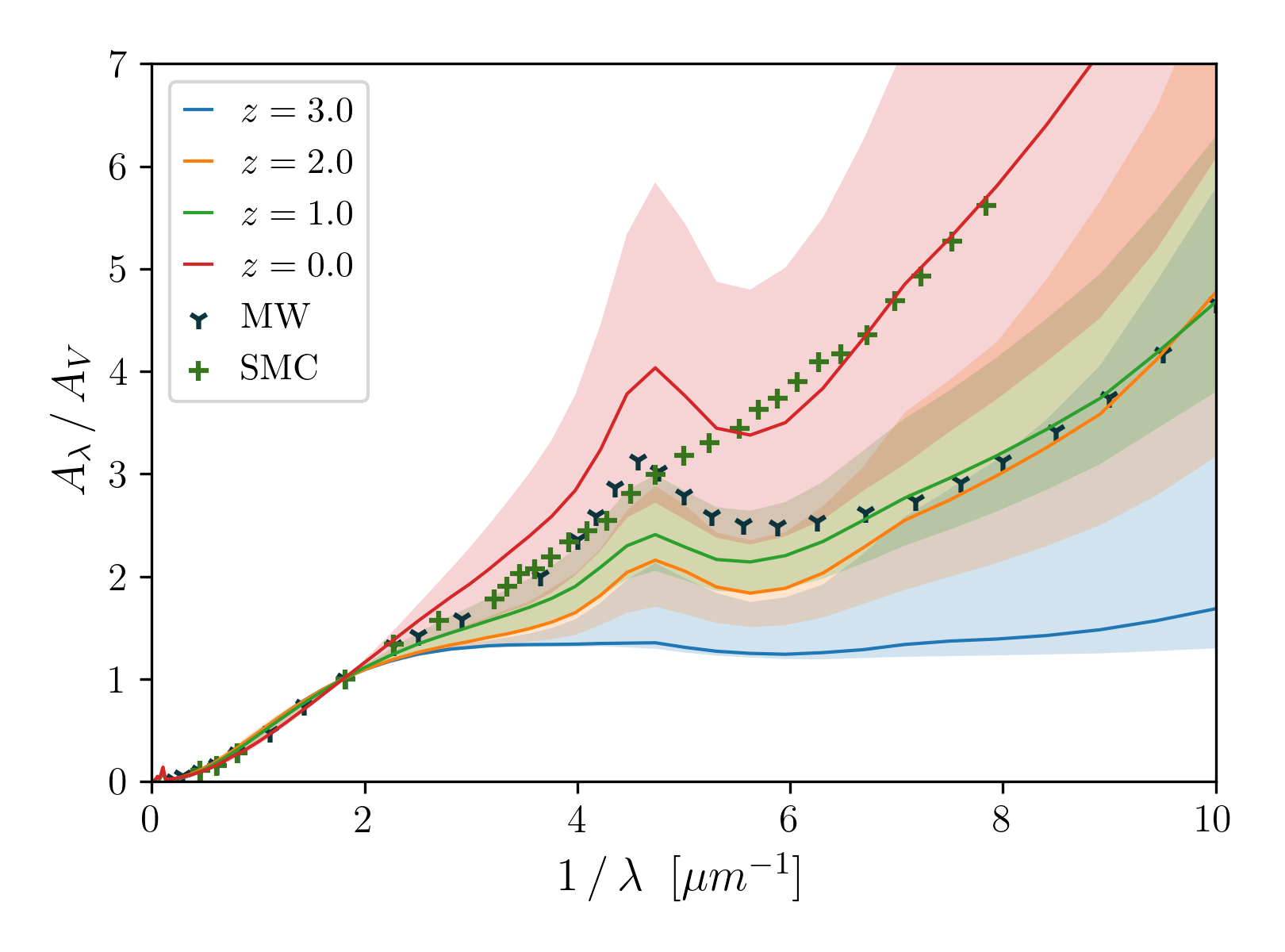}
    \caption{Extinction curves for our full MW-like galaxy sample. The blue, yellow, green, and red lines are medians at $z=3$, 2, 1, and 0, respectively. The shaded area shows the range between the 25th and 75th percentiles. The dark blue Y and the dark green cross show  observed extinction curves for the MW and the SMC, respectively from \citet{ext_obs_Pei1992}.} \label{fig:ext_curve}
\end{figure}

Based on the grain size distributions for the full sample above, we calculate extinction curves. Fig.~\ref{fig:ext_curve} shows the median extinction curves at different epochs, together with their variation. Overall, the features in extinction curve shape can be interpreted in the context of the corresponding grain size distributions.

At $z=3$, the extinction curves are flat because of the lack of small grains. When the abundance of small grains grows at later times, the extinction curves become steeper at short wavelengths. At $z=1$, because the grain size distributions are similar to MRN, the resulting extinction curves are similar to the MW extinction curve. The 2175 \AA\ bump is smaller in our calculation than in the observed MW curve. At $z=0$, because the large-grain abundance is reduced by shattering, our model predicts extinction curves that are steeper than the MW extinction curve. A typical Milky Way galaxy in our model has a $z=0$ extinction curve whose steepness is more similar to that of the Small Magellanic Cloud (SMC) extinction curve, although our predictions show a 2175 \AA\ carbon bump, which is not present in the SMC. The scatter of predicted extinction curves is large at $z=0$, and marginally covers the MW curve in the near UV and optical.

It is clear that the assembly history of a galaxy can have a significant effect on its present-day dust properties and observables, including the extinction curve. Although we cannot reject the possibility that the MW extinction curve may not be representative for our sample, selected on stellar mass and $z=0$ star formation rate, the difference between the average prediction of our sample and the MW extinction curve at $z=0$ is worth further consideration.


\section{Discussion} \label{sec:discussion}

In this section, we discuss our numerical results and their uncertainties. We also focus on some tensions with observed extinction curves, and suggest future improvements to our technique. First, we examine the relation between dust-to-gas ratio and metallicity in Section~\ref{discussion:lowDhighZ}. We compare the grain size distribution at $z=0$ to the MRN slope in Section \ref{discussion:flatness}. We also show that some of the sample galaxies in the TNG have extinction curves similar to the observed MW curve in Section \ref{subsec:MWlike_ext}. We discuss the physical reason for the scatter in the grain size distribution in Section \ref{discussion:dispersion}. Note that the slope referred to below is that of the grain size distribution in the small-size regime ($a \sim 10^{-3}$ to $0.03\,\micron$), and that dispersion refers to the 25th to 75th percentile range of the grain size distribution. 

\subsection{Dust-to-gas ratio and metallicity: outliers and redshift evolution} \label{discussion:lowDhighZ}

As shown in Section \ref{stat:DZ}, we successfully reproduced the relation between dust-to-gas ratio and metallicity at $z\sim 0$, including the nonlinear increase at sub-solar metallicity by accretion. We discuss discrepant data points and comparisons with previous work.

Figs.~\ref{fig:dz_diagram} and \ref{fig:separateDZ}d show that there are some galaxies at $z\sim 0$ located at the bottom-middle region in the $D_\mathrm{tot}$--$Z$ diagram. These galaxies have extremely low dust-to-gas ratios in spite of high ($\sim$solar) metallicity, and are preferentially found at low redshift. Since this phenomenon is not seen in one-zone models (HM20) or the simulation of an isolated galaxy by \citet{Aoyama:2017mnras}, this is likely associated with the build-up (i.e.\ merging and mass accretion) of galaxies. However, we do not see such extremely low-${D}_\mathrm{tot}$ galaxies at high metallicity in previous cosmological simulations that included similar dust evolution models \citep{Aoyama:2018mnras}. Thus, we suspect that those outliers are produced by our prescription for merging or mass accretion.

To examine the origin of these low-$D_\mathrm{tot}$ objects at solar metallicity, we inspect individual mass-growth histories. We find that these galaxies tend to have progenitors devoid of gas for a few hundred megayears. In our dust evolution model, the dust is assumed to be associated with the gas (recall that $M_\mathrm{dust}=D_\mathrm{tot}M_\mathrm{gas}$). Dust therefore disappears when all the gas is lost from the progenitor. By construction, newly accreted gas, which did not belong to any progenitor, is assumed to contain no dust, only metals. This assumption is an extreme, in the sense that we implicitly assume that dust is completely destroyed in the CGM that will be accreted onto the galaxy. As we show later, metallicities in general decline from $z\sim 1$ to $z\sim 0$, supporting the idea of fresh gas accretion from low-metallicity environments. No dust growth occurs by accretion if the gas metallicity decreases; this further suppresses the dust-to-gas ratio. Thus, it is most likely that the above outliers are caused by our treatment of gas accretion, and only exist when the dust is completely destroyed in the CGM which is subsequently accreted onto the central galaxy.

Such low-$D_\mathrm{tot}$ objects at solar metallicity were rare in the cosmological simulations analyzed by \citet{Aoyama:2018mnras} and \citet{Hou:2019aa}, who implemented dust evolution on-the-fly, albeit with a simplified treatment for the grain size distribution. We speculate that if we were to directly trace the dust content in the CGM as done in those works, we will find fewer outliers. However, even in their simulations, there are a few galaxies with particularly low dust-to-gas ratios. Thus, extreme dilution could be a real physical phenomenon that affects a small fraction of solar-metallicity galaxies, contributing to the global scatter.
 
\subsection{Grain size distributions and extinction curves} \label{discussion:flatness}

The evolution of the grain size distribution presented in Section \ref{stat:GSD} is quite similar to our previous one-zone model (HM20) and our simulation of an isolated galaxy \citep{Aoyama:2020mnras}. In particular, the following sequence of events is common: evolution begins with a large-grain dominated phase (the grain abundance is dominated by stellar dust production), undergoes a rapid increase of small grains by the accretion of gas-phase metals, and finally converges to a MRN-like shape due to coagulation. However, in this work we find that this progression breaks down from $z\sim 1$ to $z\sim 0$, where small grains begin to dominate again (Fig.\ \ref{fig:size_dist}). As a consequence, the extinction curves predicted at $z\sim 0$ are steeper than the MW extinction curve in the far-UV (Fig.\ \ref{fig:ext_curve}). 

We emphasize that the scatter of the extinction curves at $z=0$ is large enough to include the observed MW extinction curve at $1/\lambda\lesssim 6~\micron^{-1}$. The large extinctions we predict in the far-UV, however, are not consistent with the MW extinction curve. The discrepancy occurs at $\lambda\lesssim 0.16~\micron$, where the extinction curve shape is broadly governed by the grains with $a\sim\lambda /(2\upi )\lesssim 0.03~\micron$. Therefore, the high abundance ratio of small ($a<0.03~\micron$) to large ($a>0.03~\micron$) grains is the cause of the discrepancy. In our model, the relative abundance of small and large grains at late times (e.g.\ $z\sim 0$) is governed by the relative efficiencies of coagulation and shattering. Indeed, we have already shown in Section \ref{results:parameter} that the grain size distributions at later times strongly depend on the dense gas fraction ($\eta_\mathrm{dense}$) and the efficiencies of scattering and coagulation ($\omega_\mathrm{shat}$ and $\omega_\mathrm{coag}$). As shown in Section \ref{param:shat_coag} (especially Figs.\ \ref{fig:coag_results} and \ref{fig:shat_results}), the tendency towards fewer large grains at $z=0$ compared to $z=1$ is robust against changes of the coagulation and shattering efficiencies.

\begin{figure}
    \centering
    \includegraphics[width=\columnwidth]{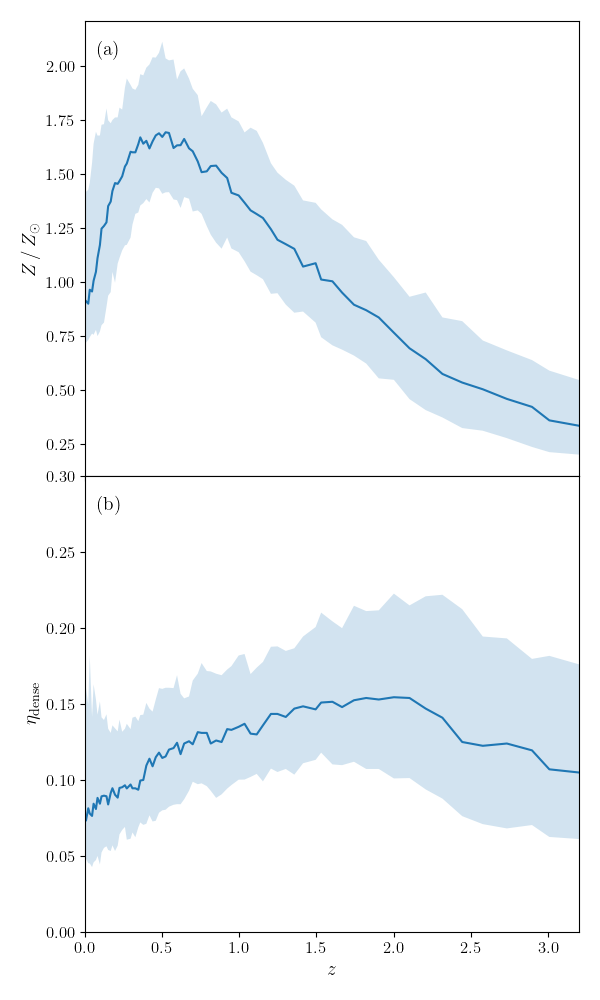}
    \caption{Evolution of the metallicity and the dense gas fraction (top and bottom panel, respectively) for the main-branch progenitors of the 206 MW-like galaxies from $z=3$ to $z=0$. In both panels, the solid curve is the median and the shaded area is the region between the 25th and 75th percentiles.}
    \label{fig:Z_evolution}
\end{figure}

Among the quantities that affect dust evolution, the dense gas fraction ($\eta_\mathrm{dense}$) and the metallicity ($Z$) vary with redshift. Fig.\ \ref{fig:Z_evolution} therefore shows the evolution of these two quantities across our sample. We observe a slight decrease in both $\eta_\mathrm{dense}$ and $Z$ from $z\sim 1$ to $z\sim 0$. The decrease of $\eta_\mathrm{dense}$ leads directly to less efficient coagulation, while the decrease of $Z$ indirectly reduces the efficiency of coagulation because of the lower dust abundance (recall that our model assumes that dust mass growth by accretion stops whenever metallicity decreases). This also leads to relatively prominent shattering. These effects that reduce coagulation largely explain the suppression of large grains from $z\sim 1$ to $z\sim 0$. 

It is worth noting that, in our fiducial model, the dense gas fraction is linked to the SFR (equation \ref{eq:KSlaw}). This means that coagulation is linked to star formation activity in our model. Thus, if the SFR declines on average from $z=1$ to $z=0$, comparable to the decline of cosmic star formation rate density at low redshift \citep{Madau:2014araa}, the efficiency of coagulation also declines. Therefore, a suppression of large grains will appear in the statistical properties of our MW-like galaxy sample.

\begin{figure}
    \centering
    \includegraphics[width=\columnwidth]{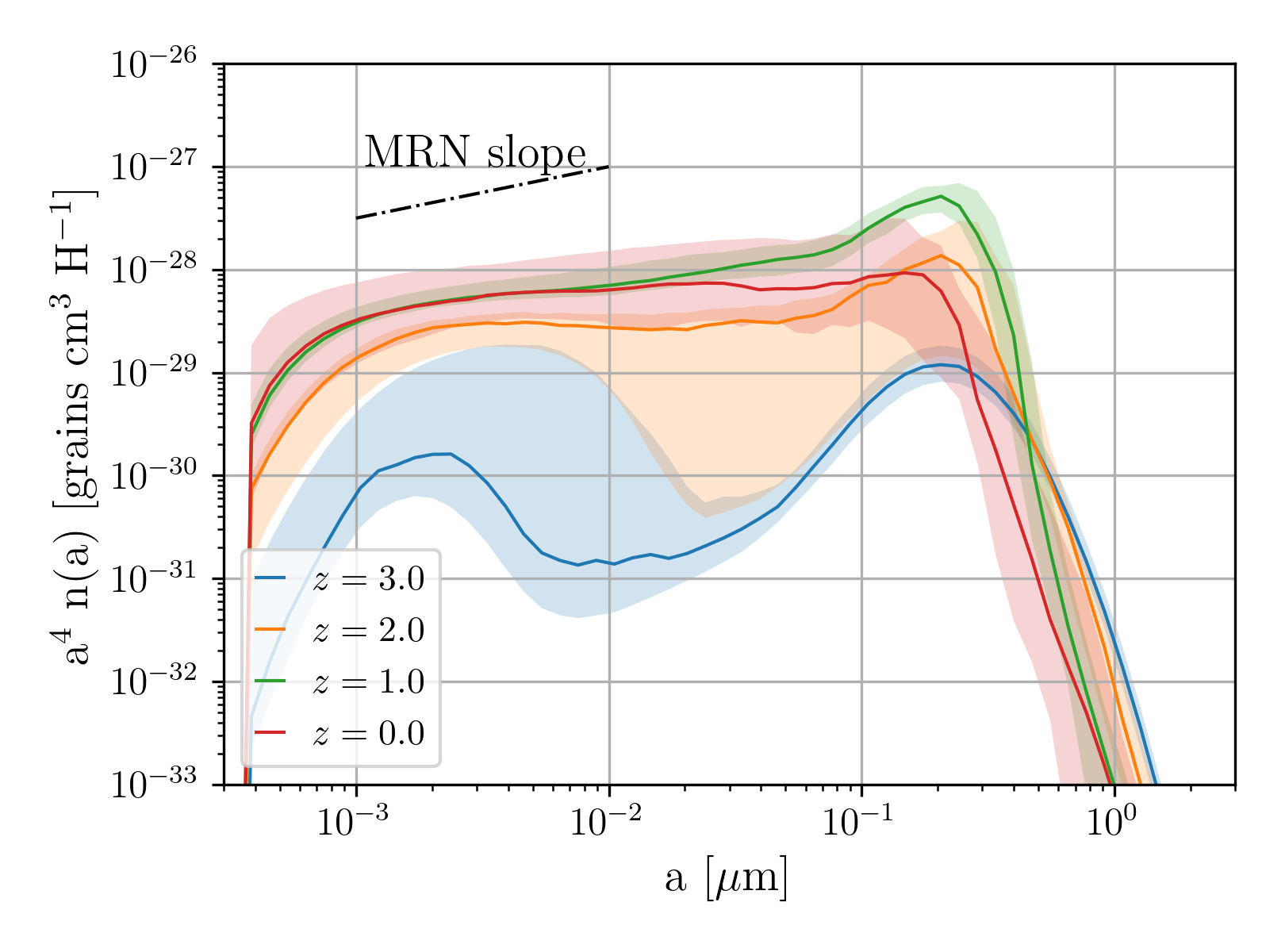}
    \caption{Grain size distributions of the 49 satellite galaxies in our sample, and their progenitors at earlier epochs. The blue, yellow, green, and red lines show the median at $z=3$, 2, 1, and 0, respectively. Shaded areas with the same colors show the corresponding 25th to 75th percentile ranges. The dot--dashed line shows the slope of the MRN grain size distribution.}
    \label{fig:satellite_size}
\end{figure}

Our selection criteria in Section \ref{model:TNG} do not take into account whether a galaxy is the central galaxy of its dark matter halo or a satellite. Based on the TNG classification, 157 out of our total of 206 galaxies are indeed centrals, and the remainder are satellites in more massive hosts. This means that Fig.~\ref{fig:size_dist} is dominated by the behaviour of central galaxies. In order to investigate the potential difference between central and satellite grain size distributions, we show the grain size distribution of the 49 satellites in Fig.~\ref{fig:satellite_size}. We do not find any significant difference, implying that the grain size distributions as well as the extinction curves are not strongly affected by whether the galaxy is a central or satellite.

\subsection{MW-like extinction curves}\label{subsec:MWlike_ext}

The observed Milky Way extinction curve does not lie close to the median of our sample at $z=0$. However, it does fall mostly within the scatter of our predictions, suggesting that some galaxies in our sample may have similar extinction curves. To locate best matches, we first quantify the similarity of each predicted extinction curve to that of the Milky Way. For this purpose, we calculate the mean squared error (MSE) between the calculated extinction curve and the MW observations \citep{ext_obs_Pei1992} as
\begin{equation} \label{eq:MSE}
    \mathrm{MSE} = \frac{1}{N}\sum_{i=1}^{N} \left[ \left( \frac{A_{\lambda_i}}{A_{V}} \right)_\mathrm{mod} - \left( \frac{A_{\lambda_i}}{A_{V}} \right)_\mathrm{obs} \right]^2,
\end{equation}
where $i$ denotes the $i$th data point, $N$ is the number of all data points used in this calculation, and the subscripts `mod' and `obs' indicate the calculated (model) and observational data, respectively. A similar criterion is used to choose MW-like extinction curves by \citet{Hou:2016pasj}. Since the predictions of our model are always quite consistent with the observations at $1/\lambda<2\,\micron^{-1}$, we restrict the MSE calculation to $1/\lambda>2\,\micron^{-1}$. We therefore include a total of 22 data points in the MSE calculation for each galaxy.

\begin{figure}
    \centering
    \includegraphics[width=\columnwidth]{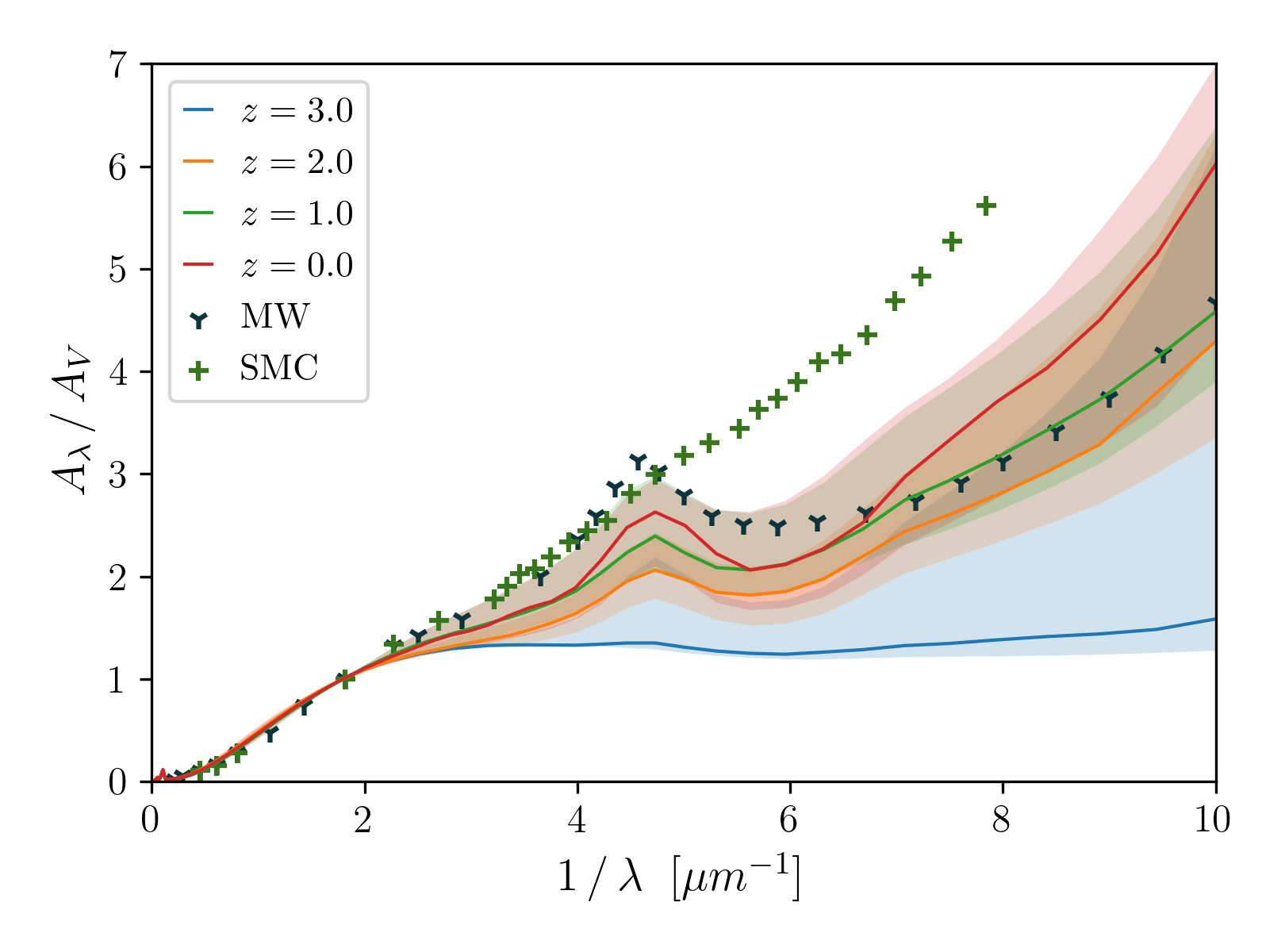}
    \caption{Same as Fig.\ \ref{fig:ext_curve} but for the subset of galaxies with extinction curves similar to the observed MW curve at $z=0$, according to the criterion described in Section \ref{subsec:MWlike_ext}.} \label{fig:MWlike_ext_stat}
\end{figure}

We select the most MW-like extinction curves as having $\mathrm{MSE}<2$. This choice is arbitrary, but the results below are insensitive to this threshold. We only include central galaxies in this comparison as the Milky Way itself is the most massive galaxy in its halo. Our criterion yields 55 galaxies out of the 158 centrals in our sample ($\sim$\,35\%) with mean halo mass $M_{200}=2.37 \times 10^{12} \mathrm{M_\odot}$ \citep[roughly consistent with empirical constraints, e.g.][]{Boylan-Kolchin:2013apj, McMillan:2017mnras}. Fig.~\ref{fig:MWlike_ext_stat} shows the median and dispersion of our MW-like extinction curves at $z=0$, together with the extinction curves of their progenitors at $z=1$--$3$. The overall agreement is significantly improved, while the slope at short wavelengths is still somewhat steeper, and the 2175 \AA\ peak is slightly less pronounced relative to the Milky Way.

The evolutionary trends of these galaxies in SFR, dense gas fraction, gas mass, and metallicity reveal why, in our model, these galaxies display MW-like extinction curves. We find that this subset has on average two times lower gas masses at $z\sim0$, at roughly the same stellar mass. This implies that $\eta_\mathrm{dense}$, according to equation (\ref{eq:KSlaw}), is relatively higher in these systems at low redshift. The dense gas not only supports a higher dust abundance through dust growth by accretion, but also leads to efficient coagulation in these galaxies. Thus, the ratio of small grains to large grains is lower and the UV slope of the extinction curve is flatter. This confirms the importance of coagulation in reproducing the MW extinction curve, as already pointed out using a one-zone model by HM20. Note that the MW sits roughly on the median $M_{\rm HI}-M_*$ and $M_{\rm HI}-\rm{SFR}$ relations ($M_\mathrm{HI}$ is the H \textsc{i} gas mass) \citep{Catinella18}, thus the high dense gas fraction may result from its relative compactness and hence high surface density.

\subsection{Dispersion in the grain size distributions} \label{discussion:dispersion}

As shown in Fig.~\ref{fig:size_dist}, the scatter among the grain size distributions in our sample of MW-like galaxies is smallest at $z=1$. This scatter could arise from galaxy-to-galaxy variations in the key quantities that determine the grain size distribution: metallicity and the dense-gas fraction. As shown in Fig.\ \ref{fig:Z_evolution}, the scatter in metallicity is approximately constant at \mbox{$\sim$\,0.5 Z$_{\sun}$}. Although $\eta_\mathrm{dense}$ has somewhat smaller scatter from $z=0.5$ to 1, there is no large variation at any redshift. Therefore, the small dispersion in the grain size distribution at $z=1$ (or the change in scatter with redshift) is not driven by variation in metallicity or dense-gas fraction.

It is therefore likely that the variation in the grain size distributions is the result of the dust evolution processes themselves. We observe in Fig.\ \ref{fig:size_dist} that the largest variation is seen at small ($a\sim 10^{-3}$--$10^{-2}~\micron$) radii for the grain size distributions at $z=3$. Since the increase in small grains is driven by accretion, the large scatter at small grain radii indicates that this process is actively ongoing for the MW progenitors at $z\sim 3$. At $z\sim 2$, there is a wide variety in the grain size distributions at $a\sim 10^{-2}$--$10^{-1}~\micron$, which reflects ongoing coagulation.

At $z\sim 1$, we observe that the grain size distributions have converged to a shape similar to the MRN distribution. In our previous one-zone calculations (HM20), it is shown that the grain size distribution roughly converges to the MRN-like shape at an age of 3--10 Gyr. In such a convergent situation, the dispersion of grain size distributions is not caused by the shape variation but by the overall dust abundance. Fig.~\ref{fig:Z_evolution} shows that the dispersion in metallicity is small ($\sim 0.2$ in $\log Z$), which explains the small dispersion in grain size distribution at $z\sim 1$. From $z=1$ to $z=0$, the dispersion of the grain size distribution becomes larger, especially at $a\gtrsim 0.1~\micron$. At this epoch, convergence to the MRN-like shape is broken by a significant drop in coagulation efficiency (compared with the shattering efficiency) for reasons given in Section \ref{discussion:flatness}. In this case, shattering becomes relatively prominent and starts to drive down the number of grains at $a\gtrsim 0.1~\micron$.

\subsection{Other simulations of the grain size distribution}

Although there are many single-galaxy or cosmological simulations that include dust evolution, only a small number of these include an evolving grain size distribution. \citet{Aoyama:2020mnras} implemented a 32-bin size distribution model in their simulation of an isolated galaxy and solved dust evolution on-the-fly, within the hydrodynamic calculation. They clarified the importance of spatial and temporal resolution, as follows. They showed that the grain size distributions are different between the dense and diffuse ISM. Thus, it is important to consider the local hydrodynamic state. They also argued that, compared to the post-processing of an isolated galaxy by \citet{2019DustModel}, the variation of the gas state (especially density) on a short time-scale has a stronger effect on the dust evolution. We note that our results at $z=0$ are intermediate between those for the dense and diffuse ISM in \citet{Aoyama:2020mnras}.

In this sense, our post-processing one-zone treatment of individual galaxies seems to capture the essence of that model; however, considering that the MW extinction curve is not reproduced on average by our model, the ratio between dense and diffuse ISM may not be correctly assigned. We should keep in mind, though, that the simulations in  \citet{Aoyama:2020mnras} also relied on a subgrid model for accretion and coagulation, because the dense clouds were not well resolved even in their isolated-galaxy simulation. Even within the post-processing framework, future work could extend our model from the single-zone assumption and instead treat dust physics at the resolution scale of individual gas cells.

Forgoing the full grain size distribution, \citet{Aoyama:2018mnras} and \citet{Hou:2019aa} adopted a two-size approximation, in which the entire grain population is divided into small and large grains, roughly separated at $a\sim 0.03~\micron$. \citet{Hou:2019aa} found that they could roughly reproduce the MW extinction curve at $z=0$, and showed that the relative abundance of small to large grains is slightly smaller at $z=1$ versus $z=0$ near solar metallicity \citep[][fig.~8]{Hou:2019aa}. Hence, their extinction curve is flatter at $z=1$ than at $z=0$. This is what we have observed in the grain size distribution in Fig.~\ref{fig:size_dist} and the extinction curves in Fig.~\ref{fig:ext_curve}. 

As before, however, the simulation of \citet{Hou:2019aa} still relies on a sub-grid treatment for the dense gas fraction. It is also interesting that the ratio between small- and large-grain abundances calculated by \citet{Hou:2019aa} is consistent with that derived from the dust emission SEDs of nearby galaxies \citep{Relano:2020aa}. With empirical information on the grain size distribution and individual dust components, we could further calculate the IR SED by assuming a typical interstellar radiation field as heating source, which would enable a future comparison between our model results and observations.
 
 
\section{Conclusions} \label{sec:conclusion}

To understand the evolution of dust grain sizes in a statistical sample of MW-like galaxies we post-process a large sample of ($\sim 200$) galaxies selected from the state-of-the-art cosmological hydrodynamical simulation TNG300, part of the IllustrisTNG project. We use a resolved, 64-bin grain size model to calculate the evolution of the grain size distribution following the method developed by HM20. We account for dust production in stellar ejecta and dust destruction by supernova shocks, based on the metal enrichment and supernova rate from the simulation. Our model takes into account interstellar processing mechanisms in two distinct ISM phases: dust growth by accretion and coagulation in the cold dense gas, and shattering in the diffuse ISM. 

Our previous models \citep{Hou:2019aa, Aoyama:2018mnras, Aoyama:2020mnras} either focused on isolated galaxies or approximated the grain size distribution with only two representative grain radii. The new 64-bin model we have described here provides a more comprehensive treatment of size-dependent dust physics, and is the first implementation of such a model in a cosmological simulation. 

We first examine our results in comparison to the observed relation between the dust-to-gas ratio and metallicity ($D_\mathrm{tot}$--$Z$ relation). We find a broad match with observational data; in particular, the nonlinear increase of the dust-to-gas ratio with increasing metallicity at sub-solar metallicity is well reproduced. We also predict that the relation (i.e. evolutionary tracks) on the $D_\mathrm{tot}$--$Z$ diagram is invariant with redshift for MW-like galaxies, except that the dust-to-gas ratio and metallicity gradually increase as the result of dust/metal enrichment.

The main focus of this work is on the grain size distributions and the extinction curves of MW-like galaxies. Our fiducial model predicts the following evolution. At $z=3$, the grain size distribution is governed by stellar dust production, which gives rise to a log-normal distribution centred at $a\sim0.1\,\micron$. We also observe a smaller peak at $a\sim0.001\,\micron$, which is created by shattering and accretion. From $z=3$ to $z=2$, the abundance of small grains at $a\sim10^{-3}$--$10^{-2}\,\micron$ grows and a maximum at $a\sim 0.2\,\micron$ develops due to coagulation at $z \sim 2$. The grain size distributions converge approximately to the MRN distribution at $z=1$, on average. Subsequently, an excess of small grains (or a depletion of large grains) appears. The relatively low abundance of large grains at $z=0$ arises from the evolution of the dense gas fraction and metallicity, both of which show a decreasing trend at $z<1$.

The evolution of the extinction curve reflects the evolving grain size distribution. Extinction curves are flat at $z=3$ because large grains dominate. As small grains become more abundant at lower redshift, extinction curves become steeper and increasingly MW-like at $z=2$ and $z=1$. At $z=0$, the median extinction curve is steeper than the MW extinction curve at $1/\lambda \gtrsim 6\,\micron^{-1}$. At longer wavelengths, the large scatter in our results for different MW analogues broadly covers the observed MW extinction curve.

The dispersion in the resulting grain size distributions is minimized near $z=1$ and is larger at lower and higher redshifts. Because the dispersions of metallicity and dense gas fraction both show little evolution from $z=3$ to 0, we conclude that the strongly evolving dispersion in the grain size distribution is driven by variations in grain processing among galaxies. In particular, the small scatter in the grain size distribution at $z=1$ is mainly due to convergence to the MRN-like grain size distribution in the presence of strong coagulation.

In summary, our model coupled to the TNG300 simulation successfully reproduces the dust-to-gas ratio vs.\ metallicity relation for nearby galaxies. The resulting extinction curves at $z\lesssim 1$ are broadly consistent with that of the MW, especially at $1/\lambda\lesssim 6~\micron^{-1}$. On average the model predicts steeper far-UV extinction curves at $z=0$ compared to the MW, although the galaxy-to-galaxy scatter in our predicted extinction curves is large.

Finally, we note that the treatment of dust coagulation and accretion relies on sub-grid modelling, not only in this work but also in other galaxy-scale simulations. Thus, our results will be improved by leveraging simulations with higher spatial resolution in the future.

\section*{Data Availability}

Data directly related to this publication and its figures are available on request from the corresponding author. The IllustrisTNG simulations are publicly available and accessible at \url{www.tng-project.org/data} \citep{Nelson19}.

\section*{Acknowledgements}

We are grateful to G. L. Granato, the referee, for useful comments.
HH thanks the Ministry of Science and Technology (MOST) for support through grant 107-2923-M-001-003-MY3 (RFBR 18-52-52006) and 108-2112-M-001-007-MY3, and the Academia Sinica for Investigator Award AS-IA-109-M02. APC is supported by the Taiwan Ministry of Education Yushan Fellowship and MOST grant 109-2112-M-007-011-MY3.
YTL is grateful for supports from MOST 108-2112-M-001-011 and MOST 109-2112-M-001-005 and a Career Development Award from Academia Sinica (AS-CDA-106-M01).


\bibliographystyle{mnras}

\appendix
\section{Resolution test on TNG100-1} \label{appendixA}

Although we treat each subhalo as a one-zone object, the mass and spatial resolution of the cosmological simulation could still impact the outcome of our dust post-processing model indirectly. To test the effects of resolution, we apply our fiducial dust evolution calculation to a higher resolution cosmological simulation, TNG100-1. Based on the same criteria as described in Section~\ref{model:TNG}, we select 18 MW-like galaxies from TNG100-1. We further constrain the sample to central galaxies only, leaving 14 galaxies in our TNG100-1 sample. 

\begin{figure}
    \centering
    \includegraphics[width=\columnwidth]{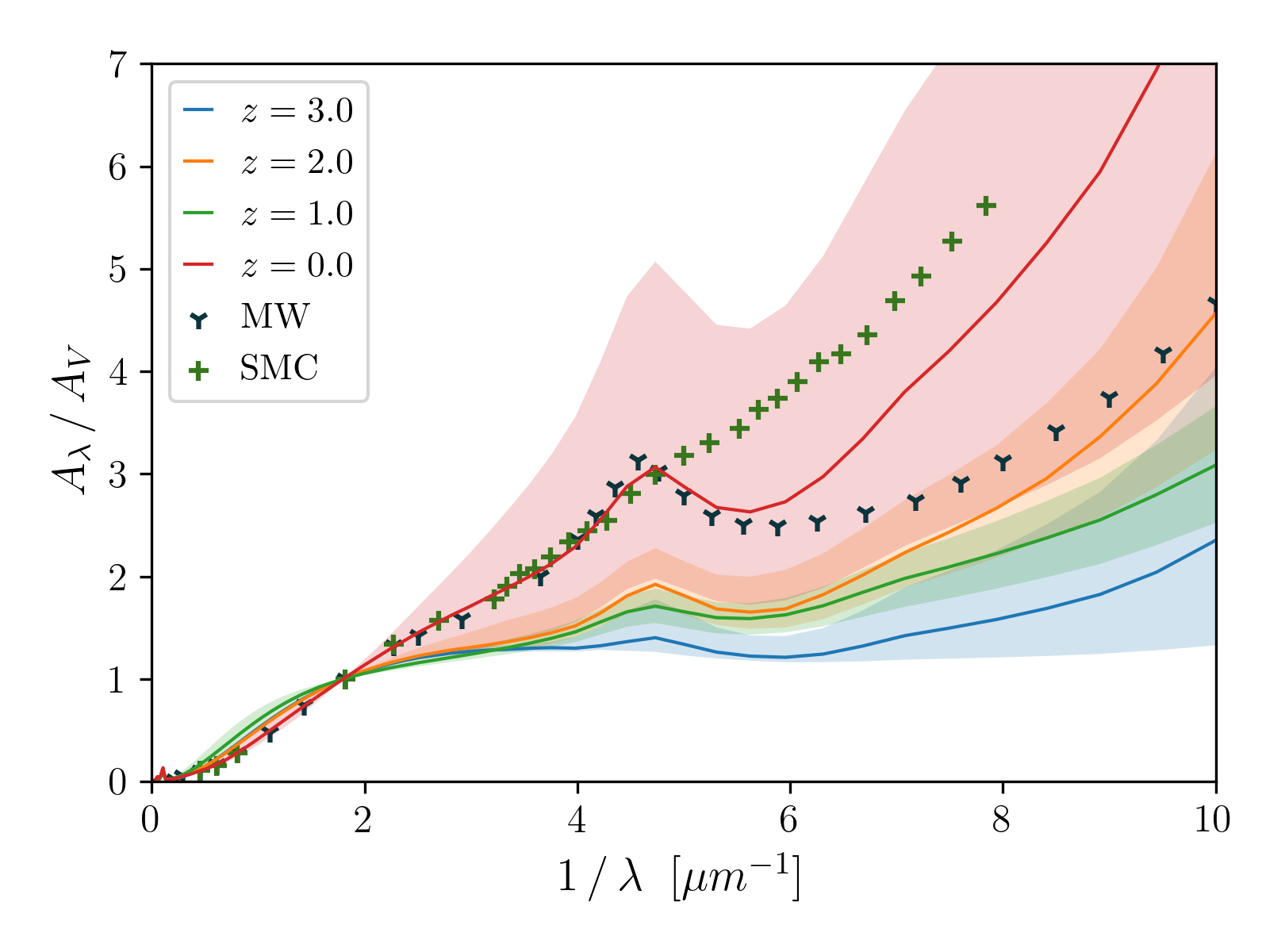}
    \caption{Extinction curves for the MW analogues in TNG100-1. The blue, yellow, green, and red lines are the medians at $z=3$, 2, 1, and 0, respectively. The shaded area shows the range between the 25th and the 75th percentiles. The dark blue Y symbols and dark green crosses show the observed extinction curves for the MW and the SMC, respectively.}
    \label{fig:tng100_ext}
\end{figure}

We run our fiducial dust model and show the resulting extinction curves in Fig.~\ref{fig:tng100_ext}. Compared with Fig.~\ref{fig:ext_curve}, the median extinction curve at $z=0$ in Fig.~\ref{fig:tng100_ext} is much more similar to the observations. The extinction curves at $z=0$ are almost consistent with the observed MW curve within the scatter of the predictions. The UV slope of the median remains steeper, implying that the average small-to-large grain ratio is still larger than that inferred for the MW.

To explain the difference between the TNG100-1 and TNG300-1 result we examine the redshift evolution of metallicity, gas mass and SFR. On average, TNG100-1 galaxies have slightly lower gas mass, lower gas-phase metallicity, and higher SFR at late times (not shown). Combined with our model for $\eta_\mathrm{dense}$ (equation \ref{eq:KSlaw}), the opposite trend of gas mass and SFR with respect to resolution results in higher $\eta_\mathrm{dense}$ in TNG100-1. Coagulation
is more efficient because of higher $\eta_\mathrm{dense}$; hence, more large grains are able to form and the small-to-large grain ratio is smaller. Eventually, the decrease of the small-to-large grain ratio in grain size distributions leads to less steep extinction curves in Fig.~\ref{fig:tng100_ext} compared to Fig.~\ref{fig:ext_curve}.

Ultimately, both gas mass and SFR are affected by the hydrodynamical simulation and its treatment of cooling, star formation, stellar feedback and black hole feedback (active galactic nuclei). The detailed behaviour of these processes can depend in non-trivial ways on numerical resolution, as discussed in \citet[][appendix A]{Pillepich18a} and \citet[][appendix B]{Weinberger17}. In particular, \citet{Pillepich18a} demonstrates that while galactic properties such as those input into our modeling are converging, they are not yet fully converged at the resolution of TNG300-1, nor TNG100-1. Understanding and optimizing the numerical convergence of cosmological galaxy formation models remains an ongoing challenge.

Although we focused on the differences, the large dispersions in both simulations overlap significantly. The number of TNG100-1 galaxies is also too small to put strong statistical constraints on resolution effects. The qualitative evolutionary trends of the extinction curves with redshift in fact remain unchanged, and we conclude that the overall picture and predictions presented in this work are not strongly impacted by numerical resolution.

\bsp	
\label{lastpage}
\end{document}